\documentclass[aps,prl,twocolumn,floatfix]{revtex4-1}

\usepackage{graphicx}
\usepackage[utf8]{inputenc}
\usepackage{amsmath,amsfonts,amssymb}
\usepackage{color}

\begin{document}

\title{Single-Co Kondo effect in atomic Cu wires on Cu(111)}

\author{N.\ Néel}
\affiliation{Institut für Physik, Technische Universität Ilmenau, D-98693 Ilmenau, Germany}
\author{J.\ Kröger}
\affiliation{Institut für Physik, Technische Universität Ilmenau, D-98693 Ilmenau, Germany}
\author{M.\ Schüler}
\affiliation{Bremen Center for Computational Materials Science, University Bremen, D-28359 Bremen, Germany}
\affiliation{Institute for Theoretical Physics, University Bremen, D-28359 Bremen, Germany}
\author{B.\ Shao}
\affiliation{Bremen Center for Computational Materials Science, University Bremen, D-28359 Bremen, Germany}
\affiliation{Institute for Theoretical Physics, University Bremen, D-28359 Bremen, Germany}
\author{T. O. Wehling}
\affiliation{Bremen Center for Computational Materials Science, University Bremen, D-28359 Bremen, Germany}
\affiliation{Institute for Theoretical Physics, University Bremen, D-28359 Bremen, Germany}
\author{A.\ Kowalski}
\affiliation{Institut für Theoretische Physik und Astrophysik and Würzburg-Dresden Cluster of Excellence ct.qmat, Universität Würzburg, D-97074 Würzburg, Germany}
\author{G.\ Sangiovanni}
\affiliation{Institut für Theoretische Physik und Astrophysik and Würzburg-Dresden Cluster of Excellence ct.qmat, Universität Würzburg, D-97074 Würzburg, Germany}

\begin{abstract}
Linear atomic chains containing a single Kondo atom, Co, and several nonmagnetic atoms, Cu, were assembled atom by atom on Cu(111) with the tip of a scanning tunneling microscope.
The resulting one-dimensional wires, Cu$_m$CoCu$_n$ ($0\leq m, n\leq 5$), exhibit a rich evolution of the single-Co Kondo effect with the variation of $m$ and $n$, as inferred from changes in the line shape of the Abrikosov-Suhl-Kondo resonance.
The most striking result is the quenching of the resonance in CuCoCu$_2$ and Cu$_2$CoCu$_2$ clusters.
State-of-the-art first-principles calculations were performed to unravel possible microscopic origins of the remarkable experimental observations.  
\end{abstract}

\maketitle

\section{Introduction}

The Kondo effect was originally observed as the resistivity increase below a characteristic temperature of some metals containing magnetic impurities. \cite{annphys_7_761,physica_34_1115}
As intuited by J.\ Kondo in 1964, many-body correlations are at the origin of this phenomenon. \cite{ptp_32_37,pr_169_437}
Later, his ideas were extended beyond the perimeter of perturbation theory by P.\ W.\ Anderson \cite{pr_124_41,jpcssp_3_2436} and by K.\ Wilson \cite{rmp_47_773} understanding and exploiting the role of scale invariance. 
The conduction electrons of the metal scatter off the local spin of a single quantum impurity and a low-energy collective excitation described by a quantum mechanical singlet wave function is formed.
The bare magnetic moment is hence screened and, oversimplifying, one can picture this mechanism as multiple spin flips occurring due to the exchange interaction between the impurity spin and the conduction electrons.

The spectroscopic signature of this spin exchange is represented by the Kondo \cite{ptp_32_37,pr_169_437} or Abrikosov-Suhl \cite{ppf_2_5,ppf_2_61,aca_1967} resonance at the Fermi energy, $E_\text{F}$\@.
At zero temperature the resonance full width at half maximum is related to the Kondo temperature $T_\text{K}$ \textit{via} $\text{k}_\text{B}T_\text{K}$ ($\text{k}_\text{B}$: Boltzmann constant)\@. 
The presence of this resonance was confirmed in photoemission \cite{prl_55_1518,prl_58_2810,prb_42_8864,prl_65_1639,prb_44_8304,prb_76_045117} and inverse photoemission \cite{prb_28_7354} experiments.

It is certainly no exaggeration to say that the detection of the Abrikosov-Suhl-Kondo (ASK) resonance at the single-impurity level with a scanning tunneling microscope (STM) \cite{prl_80_2893,science_280_567} represents a seminal finding in surface physics.
In spectra of the differential conductance ($\text{d}I/\text{d}V$, $I$: current, $V$: bias voltage) the ASK resonance appears with a Fano \cite{nc_12_154,pr_124_1866} or Frota \cite{prb_33_7871,prb_45_1096} line shape.
The Fano line shape reflects that spin-conserving tunneling occurs directly from the tip to the conduction electron continuum and indirectly \textit{via} the Kondo resonance.
The interference of tunneling electrons in these two paths leads to the characteristic spectroscopic line shape. \cite{prl_80_2893,science_280_567}
Many-body numerical renormalization group calculations \cite{prb_84_195116} justify the phenomenological Frota line shape, which was likewise reported from experiments. \cite{natphys_7_203,prl_108_166604}
A wealth of reports followed the initial observations \cite{prl_80_2893,science_280_567} and revealed the rich physics of the single-atom \cite{prl_88_077205,prl_88_096804,prl_93_176603,prl_108_266803,prl_114_076601,prb_91_201111} and single-molecule \cite{prl_95_166601,science_309_1542,prl_97_266603,nl_6_820,prl_99_106402,prl_101_217203,prl_105_106601,nl_10_4175,natcommun_2_217,natcommun_2_490,prl_106_187201,science_332_940,nl_12_3174,nl_12_3609,acie_51_6262,natcommun_3_938,prl_109_086602,acsnano_7_9312,nl_13_4840} Kondo effect.  
Excellent review articles are available, \textit{e.\,g.}, Refs.\,\onlinecite{jpcm_21_053001,ss_630_343,pss_92_83}.

Importantly, variations of the ASK resonance line shape in spectroscopy experiments with an STM serve as a subtle probe for magnetic interaction, \cite{prb_60_r8529,prb_66_212411,prl_98_056601,natphys_4_847,prl_103_107203,prl_107_106804,nl_12_3174,natphys_7_901,natcommun_5_5417,natcommun_6_10046,nl_16_6298} hybridization \cite{science_309_1542,prl_97_266603,nl_6_820,prl_98_016801,prl_101_266803,prb_82_233401,nl_14_3895,prb_91_201111,nl_17_7146} and density of states (DOS) \cite{prl_101_266803,ass_237_576,prb_78_033402,prb_97_035417} effects.
In particular, a nonmonotonic evolution of $T_\text{K}$ of a single Co atom in a compact CoCu$_n$ cluster on Cu(111) \cite{prl_101_266803} unraveled that the local electronic structure at the magnetic-impurity site is essential for its Kondo effect, rather than the sheer number of coordinations with nearest-neighbor atoms.
Deviating from the previous \textit{compact} cluster geometry \cite{prl_101_266803} the evolution of the Kondo effect in \textit{linear} Cu$_m$CoCu$_n$ clusters is reported here.
All atomic chains exhibit ASK resonances with $40\,\text{K}\leq T_\text{K}\leq 160\,\text{K}$ with two surprising and notable exceptions: the ASK resonances at the Co site for CuCoCu$_2$ and Cu$_2$CoCu$_2$ clusters are quenched, at odds with all other clusters that show a clear spectroscopic signature of the Kondo effect.

Motivated by these appealing experimental results, density functional theory (DFT) and quantum monte carlo (QMC) calculations were applied in order to describe the Kondo physics of the linear clusters.
While many properties of the clusters are captured by the calculations, the atomically fabricated Cu$_m$CoCu$_n$ clusters define a test case for systematically benchmarking theory against experiment.  
As a starting point towards such a benchmarking, DFT and QMC calculations were combined in a "DFT++ manner" \cite{epjst_226_2439} to test to which extent experimental findings can or cannot be explained by this approach.
Our work unveils the capabilities and open issues of state-of-the-art modeling techniques.

\section{Experiment}

Experiments were performed with an STM operated in ultrahigh vacuum ($10^{-9}\,\text{Pa}$) and at $5\,\text{K}$\@.
Cu(111) surfaces were cleaned by Ar$^+$ bombardment and annealing.
Chemically etched W wires -- presumably coated with Cu substrate material due to \textit{in situ} preparation by indentations -- served as STM tips.
Single Co atoms were deposited at $\approx 8\,\text{K}$ from an electron beam evaporator through openings of the radiation shields of the bath cryostat.
 Single Cu atoms were transferred from the tip to the surface by controlled tip-surface contacts. \cite{prl_94_126102,njp_9_153,jpcm_20_223001,njp_11_125006}
STM images were recorded at constant current with the bias voltage applied to the sample.
Spectra of $\text{d}I/\text{d}V$ were acquired by sinusoidally modulating the bias voltage ($1\,\text{mV}_{\text{rms}}$, $950\,\text{Hz}$) and measuring the first harmonic of the current response of the tunneling junction with a lock-in amplifier.
Topographic data were processed using WSXM. \cite{rsi_78_013705}

\section{Results and discussion}

\subsection{Scanning tunneling microscopy and spectrocopy}

\begin{figure}
\includegraphics[width=\columnwidth]{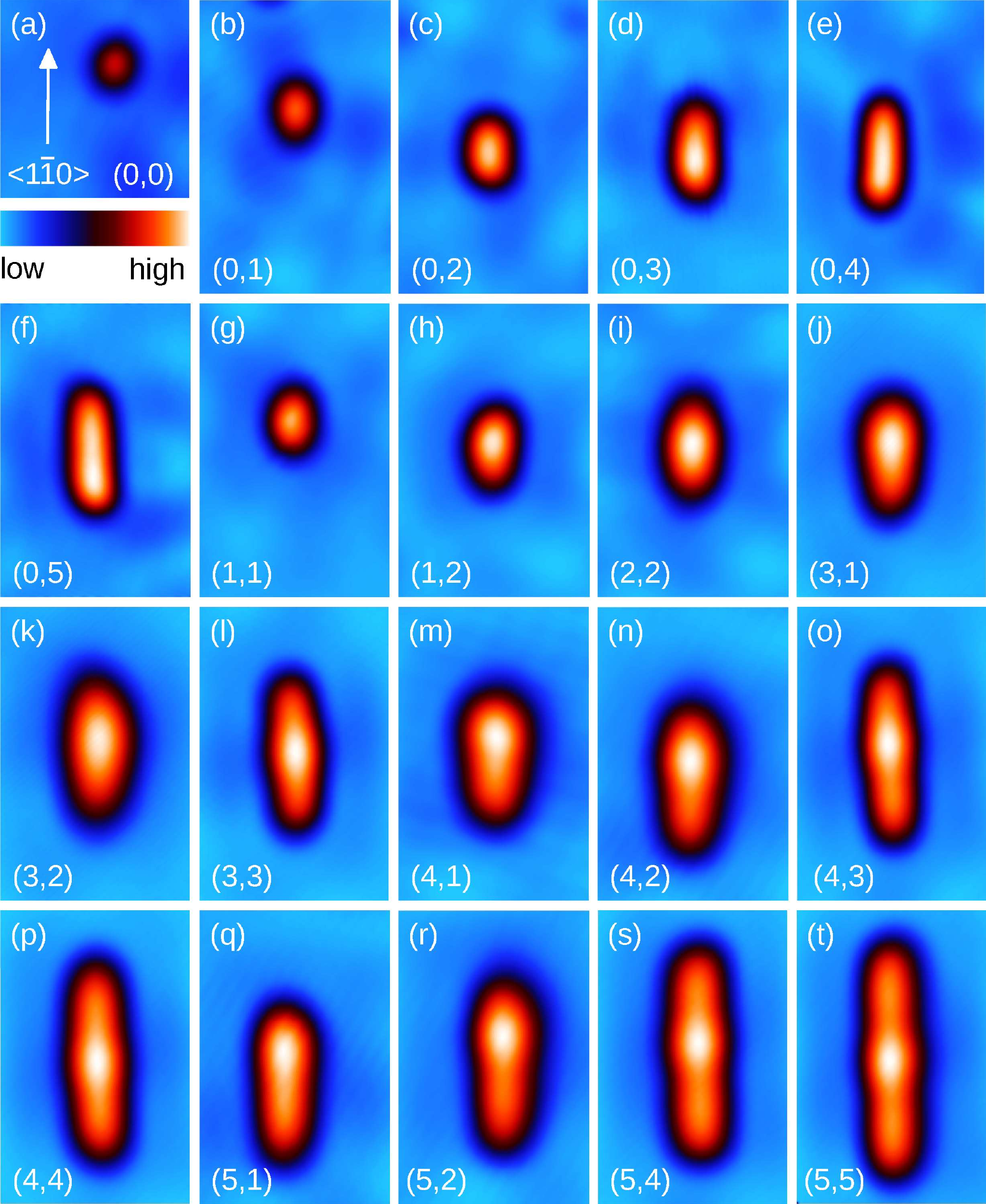}
\caption[fig1]{(Color online)
STM images of atom-by-atom fabricated linear Cu$_m$CoCu$_n$ clusters on Cu(111) denoted $(m,n)$ with $0\leq m,n\leq 5$\@ (bias voltage: $10\,\text{mV}$, tunneling current: $50\,\text{pA}$)\@.
(a) STM image of a single Co atom (size: $2.6\times 2.7\,\text{nm}^2$)\@.
A Cu(111) crystallographic direction, $\langle 1\bar{1}0\rangle$, is indicated by an arrow.
The color scale ranges from $0$ (low) to $120\,\text{pm}$ (high) in all STM images.
(b)--(t) STM images of linear clusters ($2.6\times 4\,\text{nm}^2$)\@.  
}
\label{fig1}
\end{figure}

Figure \ref{fig1} shows representative STM images of the Cu$_m$CoCu$_n$ clusters studied in this work.
In the following these assemblies will be referred to as $(m,n)$-clusters.
Single Co [Fig.\,\ref{fig1}(a)] and Cu atoms on Cu(111) appear as circular protrusions with larger apparent height for Co ($\approx 70\,\text{pm}$) compared to Cu ($\approx 50\,\text{pm}$) at $V = 10\,\text{mV}$\@.
The $(m,n)$-clusters were fabricated atom by atom. \cite{science_313_948,prl_98_146804}
After assembling the linear Cu$_m$ cluster, one edge of the chain was decorated by a single Co atom.
Additional $n$ Cu atoms were then attached to this Co atom, continuing the direction of the chain.
Attaching a single Cu atom to Co [Fig.\,\ref{fig1}(b)] increases the Co apparent height to $\approx 90\,\text{pm}$.
$(0,2)$-clusters [Fig.\,\ref{fig1}(c)] exhibit a Co apparent height of $\approx 110\,\text{pm}$.
Additional Cu atoms in $(0,n)$-clusters ($n\geq 3$) do not change the Co apparent height appreciably [Fig.\,\ref{fig1}(d)--(f)]\@.
Similar observations hold for $(m,n)$-clusters with $m\geq 1$ [Fig.\,\ref{fig1}(g)--(t)]\@.
Inside such clusters Co always appears $\approx 15\,\text{pm}$ higher than its Cu neighbors.
All linear atomic chains are aligned with $\langle 1\bar{1}0\rangle$ crystallographic directions of Cu(111) [arrow in Fig.\,\ref{fig1}(a)]\@.

\begin{figure}
\includegraphics[width=\columnwidth]{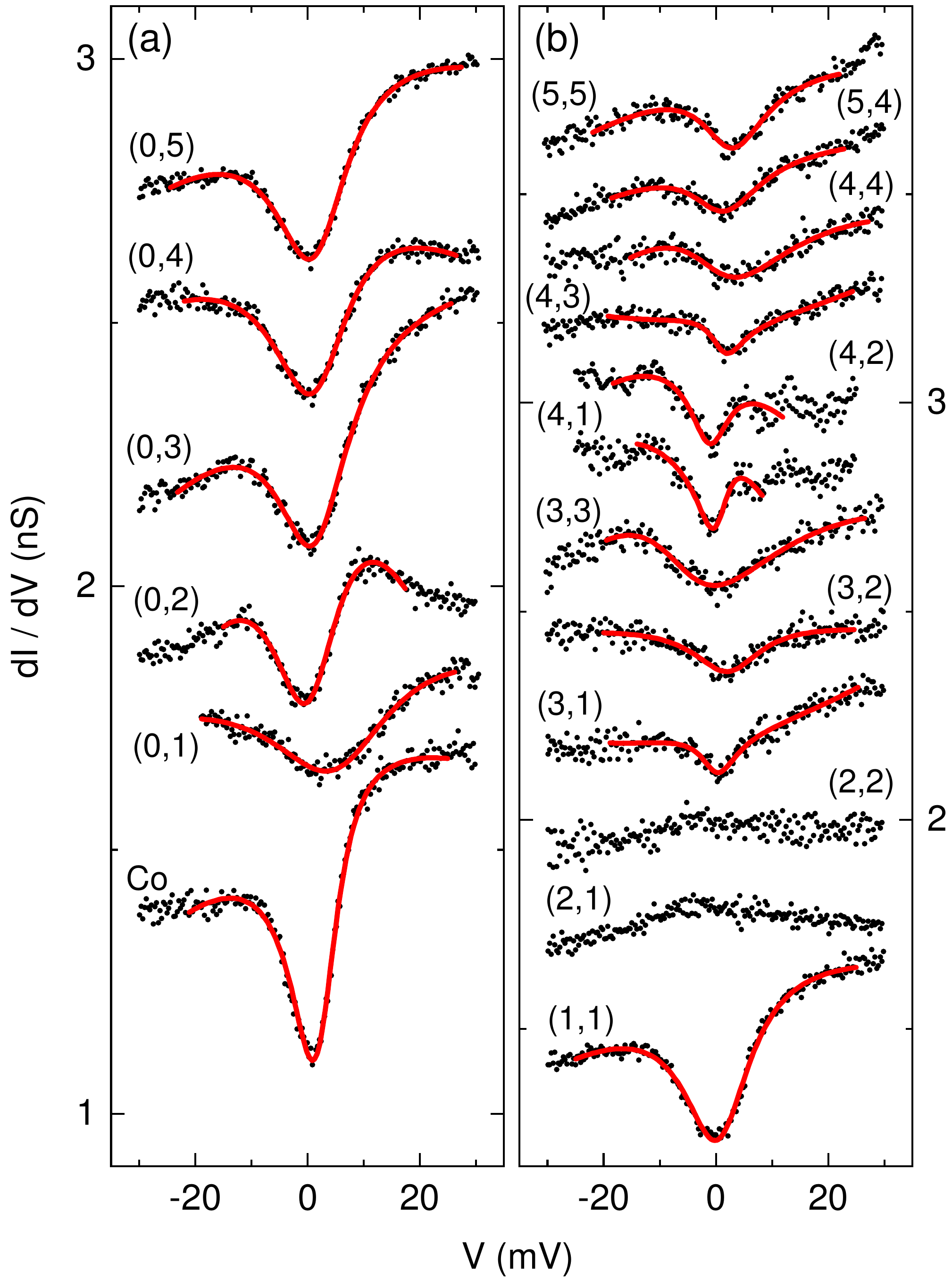}
\caption[fig2]{(Color online)
Spectroscopy of the ASK resonance.
Spectra of $\text{d}I/\text{d}V$ (dots) recorded above the Co atom in $(m,n)$-clusters on Cu(111) for (a) $m=0$, $n\geq 1$ and (b) $m,n\geq 1$\@.
The bottom spectrum of (a) depicts the single-Co ASK resonance.
Solid lines are fits to the data using the Fano line shape (see text)\@.
Spectra are offset for clarity.
Feedback loop parameters prior to data acquisition: $30\,\text{mV}$, $50\,\text{pA}$\@. 
}
\label{fig2}
\end{figure}

\begin{table}
\caption[tab1]{
Kondo temperature $T_\text{K}$ (K) of $(m,n)$-clusters extracted from fits of $\text{d}I/\text{d}V$ spectra using the Fano line shape.
}
\begin{ruledtabular}
\begin{tabular}{c|cccccc}
$(m,n)$ & $0$ & $1$ & $2$ & $3$ & $4$ & $5$ \\
\hline
$0$ & $53\pm 2$ & $153\pm 15$ & $87\pm 5$ & $90\pm 4$ & $94\pm 4$ & $91\pm 3$ \\
$1$ & $153\pm 15$ & $81\pm 3$ & / & $45\pm 3$ & $39\pm 6$ & \\
$2$ & $87\pm 5$ & / & / & $85\pm 5$ & $49\pm 8$ & \\
$3$ & $90\pm 4$ & $45\pm 3$ & $85\pm 5$ & $144\pm 13$ & $41\pm 5$ & \\
$4$ & $94\pm 4$ & $39\pm 6$ & $49\pm 8$ & $41\pm 5$ & $119\pm 10$ & $87\pm 7$ \\
$5$ & $91\pm 3$ & & & & $87\pm 7$ & $79\pm 6$ \\
\end{tabular}
\end{ruledtabular}
\label{tab1}
\end{table}

The one-dimensional wires were used to explore changes in the Co Kondo effect.
Spectra of $\text{d}I/\text{d}V$ are summarized in Fig.\,\ref{fig2}, for $(0,n)$-clusters ($0\leq n\leq 5$) in Fig.\,\ref{fig2}(a) and for $(m,n)$-clusters ($1\leq m, n\leq 5$) in Fig.\,\ref{fig2}(b)\@.
All spectra were acquired atop the Co central region.
The majority of spectra exhibits an indentation close to zero bias voltage, which is interpreted as the ASK resonance.
Solid lines are fits to the data using the Fano line shape $f(V)=a\cdot(q+\varepsilon)^2/(1+\varepsilon^2)$ [$a$: amplitude, $q$: asymmetry parameter, $\varepsilon=\beta\cdot(\text{e}V-\varepsilon_0)$ with $1/\beta=\text{k}_\text{B}T_\text{K}$, e the elementary charge, $\varepsilon_0$ the resonance energy] and a parabolic background.

The single Co atom [bottom data set of Fig.\,\ref{fig2}(a)] exhibits an ASK resonance with a width that corresponds to $T_\text{K}=53\pm 2\,\text{K}$ (see Table \ref{tab1} for a collection of all extracted $T_\text{K}$), which is in agreement with previous reports. \cite{prl_88_096804,prl_101_266803}
By adding one Cu atom [$(0,0)\rightarrow(0,1)$] $T_\text{K}$ increases by nearly a factor $3$ to $T_\text{K}=153\pm 15\,\text{K}$\@.
This increase of $T_\text{K}$ is in accordance with a previous work on compact CoCu$_n$ clusters. \cite{prl_101_266803}
The addition of a second Cu atom [$(0,1)\rightarrow(0,2)$] gives rise to a considerable decrease of $T_\text{K}$ to $87\pm 5\,\text{K}$, which for $(0,n)$-clusters ($n\geq 2$) then stays nearly constant.
$T_\text{K}$ values of $90\pm 4\,\text{K}$ for $(0,3)$ and $94\pm 4\,\text{K}$ for $(0,4)$ agree well with previous experimental observations and calculations for linear CoCu$_3$ and CoCu$_4$ clusters. \cite{prl_107_106804}

Clusters $(m,n)$ with $m,n\geq 1$ unveil unexpected behavior.
The most remarkable observation is the entire quenching of the ASK resonance of Co in $(1,2)$, $(2,1)$, $(2,2)$-clusters [second and third spectrum counted from the bottom in Fig.\,\ref{fig2}(b)]\@.
This behavior was reproduced for several clusters and tips.
The suitability of the tip for spectroscopic measurements was intermediately controlled by reproducing the single-Co ASK resonance and the steplike onset of the Cu(111) Shockley surface state (not shown)\@.
In order to exclude the presence of exceptionally wide ASK resonances \cite{prl_108_266803} which would appear rather flat on the small bias voltage range of Fig.\,\ref{fig2}, the bias voltage range in $\text{d}I/\text{d}V$ spectroscopy was extended to $\pm 500\,\text{mV}$\@.
An ASK resonance was not observed.
Moreover, for the larger clusters a maximum of $T_\text{K}$ was reached in the symmetric case, \textit{i.\,e.}, for $(3,3)$-clusters with $T_\text{K}=144\pm 13\,\text{K}$ and for $(4,4)$-clusters with $T_\text{K}=119\pm 10\,\text{K}$\@.
In both cases, the addition of a single Cu atom to one side of the linear chain leads to a significant drop of $T_\text{K}$, $T_\text{K}=41\pm 5\,\text{K}$ for $(3,4)$-clusters and $T_\text{K}=87\pm 7\,\text{K}$ for $(4,5)$-clusters.
$(5,5)$ -clusters do not adhere to this trend, which may indicate that these very long clusters behave like infinite Cu chains with an embedded Co atom.
The asymmetry parameter $q$ varied between $0.1$ and $0.2$ for all clusters without showing characteristic behavior with $m$ and $n$.

Before turning to the calculations it is important to mention that changes in $T_\text{K}$ observed from the small linear clusters reported here deviate from the previously reported evolution of $T_\text{K}$ in compact CoCu$_n$ clusters. \cite{prl_101_266803}
Indeed, for compact CoCu$_n$ clusters $T_\text{K}$ was demonstrated to first increase with $n$ up to $n=2$ with a maximum value $T_\text{K}\approx 326\,\text{K}$ and then decrease down to $T_\text{K}\approx 43\,\text{K}$ for $n=4$\@.
In contrast, the short linear clusters discussed here reach a maximum of $T_\text{K}$ for $(0,1)$\@.
Increasing the number of Cu atoms by one leads to a nearly halved $T_\text{K}$ for $(1,1)$\@.
Obviously, the Co coordination number in compact CoCu$_2$ and linear CuCoCu clusters is identical.
However, the different behavior of $T_\text{K}$ in the two cases evidences the impact of the actual cluster geometry on the Kondo effect.

\begin{figure}
\includegraphics[width=\columnwidth]{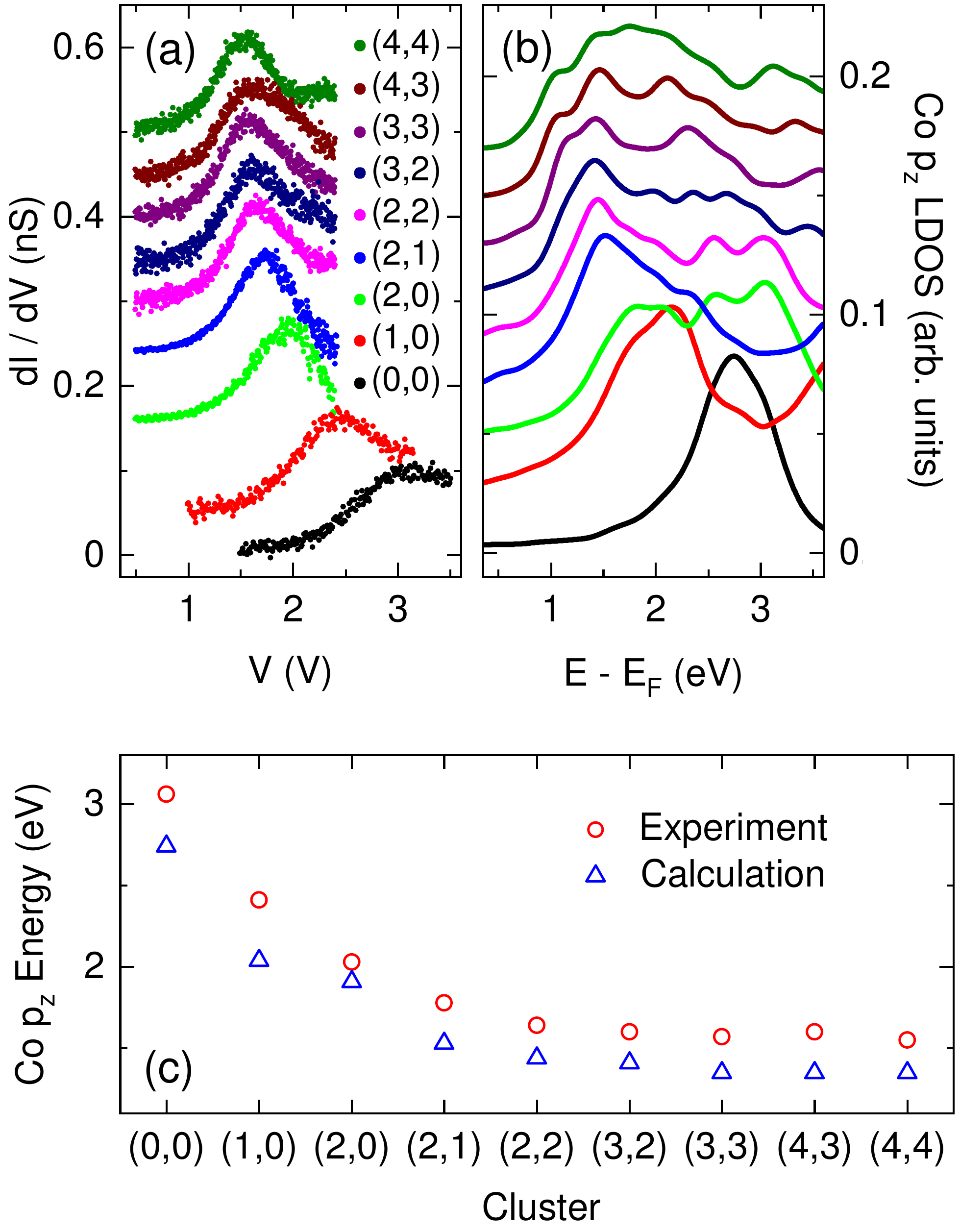}
\caption[fig3]{(Color online)
Evolution of the unoccupied Co $p_z$ resonance with the cluster size.
(a) $\text{d}I/\text{d}V$ spectra of an unoccupied state recorded above Co in $(m,n)$-clusters $(0\leq m,n\leq 4)$\@.
(b) Calculated Co $p_z$ DOS in $(m,n)$-clusters.
(c) Experimental (circles) and calculated (triangles) unoccupied resonance energy at the Co site in $(m,n)$-clusters. 
}
\label{fig3}
\end{figure}

For the calculations to be discussed next a relaxed cluster geometry is important. 
To obtain an indication for the matching between the fully relaxed geometries in theory and the experimental clusters the evolution of $\text{d}I/\text{d}V$ spectra at elevated bias voltage is compared with the calculated DOS (Fig.\,\ref{fig3})\@.
Figure \ref{fig3}(a) shows the spectroscopic signature of an unoccupied resonance above the Co site in $(m,n)$-clusters with increasing (from bottom to top) cluster length.
The resonance peak shifts to lower bias voltage with increasing $m$ and $n$.
Calculations of the local DOS at the Co site [Fig.\,\ref{fig3}(b)] reveals an unoccupied Co $p_z$ resonance whose onset likewise shifts to lower energy with increasing cluster size.
Owing to the same trend and to similar onset energies [Fig.\,\ref{fig3}(c)] the peaks in $\text{d}I/\text{d}V$ spectra are assigned to the spectroscopic signature of the calculated Co\,$p_z$ resonance.
Thus, the experimental and calculated evolutions of spectra at elevated energy are consistent, as expected for relaxed geometries.

\subsection{Model calculations}

\subsubsection{Density functional theory} 

Simulations of the Cu$_m$CoCu$_n$ chains were performed on Cu(111) slabs with a lateral extent of $3\times 13$ unit cells.
For calculations of the geometrical structure $3$ Cu layers were used, while for the hybridization simulations $5$ Cu layers were included.  
The calculations are based on the Vienna \textit{ab initio} simulation package (VASP) \cite{jpcm_6_8245} with the projector augmented wave (PAW) basis sets \cite{prb_50_17953,prb_59_1758} and the generalized gradient approximation (GGA) to the exchange correlation potential. \cite{prl_77_3865}

\begin{figure}
\includegraphics[width=\columnwidth]{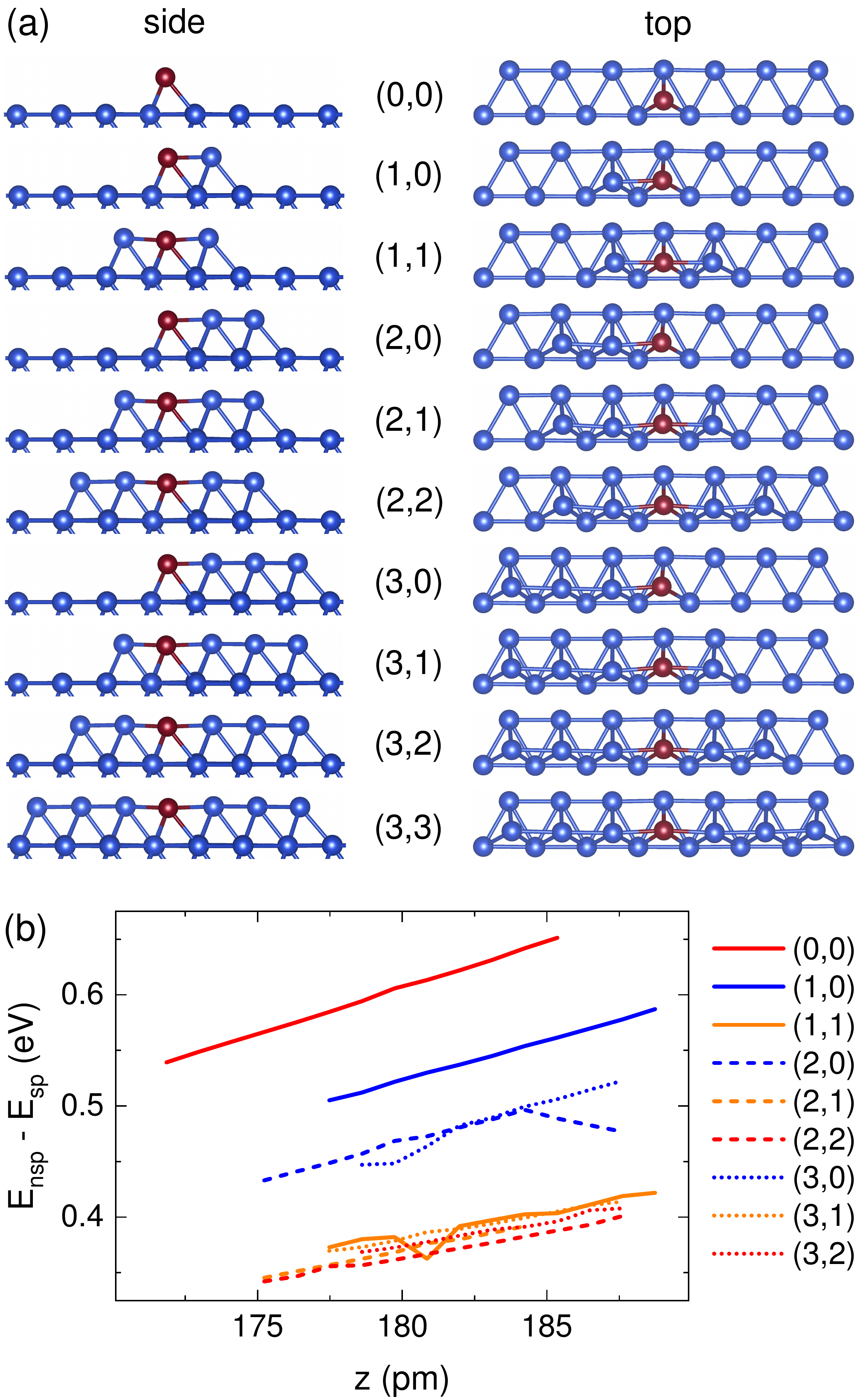} 
\caption[fig4]{(Color online) 
Simulation of $(m,n)$-cluster geometry on Cu(111) and comparison of total energies\@. 
(a) Side and top view of the relaxed chain structures as obtained from spin-polarized calculations (Cu: blue; Co: red; in the top view only Cu surface atoms closest to the linear chain are shown)\@. 
(b) Total energy differences, $E_{\text{nsp}}-E_{\text{sp}}$, of nonmagnetic and magnetic solution for fixed geometries interpolating between the non-spin-polarized (nsp) and spin-polarized (sp) structures. 
The height $z$ of the Co atom above the Cu surface is chosen as interpolation coordinate, where the lowest (highest) $z$ value for each curve corresponds to the structure obtained from the non-spin-polarized (spin-polarized) relaxation.}
\label{fig4}
\end{figure}

\textit{Atomic structure and magnetic moment formation.} 
We firstly investigated the atomic structure of the adsorbed linear $(m,n)$-clusters and its dependence on magnetic moment formation by comparing structural relaxations and total energies as resulting from non-spin-polarized (nsp) and spin-polarized (sp) calculations. 
Figure \ref{fig4}(a) shows side and top views of the relaxed structures of clusters assembled in the experiment. 
The chains are almost perfectly linear with only slight vertical ($<10\,\text{pm}$) and lateral ($<20\,\text{pm}$) corrugations.

\begin{table}
\caption{
Adsorption height of Co above Cu(111) for relaxation without (nsp), $z_{\text{Co}}^{\text{nsp}}$, and with (sp), $z_{\text{Co}}^{\text{sp}}$, spin polarization.
The magnetic moment of Co, $\mu_{\text{Co}}$, in units of the Bohr magneton $\mu_\text{B}$ originates from all Co electrons.
$E_{\text{nsp}}-E_{\text{sp}}$ denotes the total energy difference between nonmagnetic and magnetic solutions.
}
\begin{ruledtabular}
\begin{tabular}{ccccc}
$(m,n)$ & $z^\text{nsp}_\text{Co}$ (pm) & $z^\text{sp}_\text{Co}$ (pm) & $\mu_{\text{Co}}$ ($\mu_\text{B}$) & $E_\text{nsp}-E_\text{sp}$ (eV) \\ 
\colrule
$(0,0)$ & $172$ & $185$ & $1.99$ & $0.61$ \\ 
$(1,0)$ & $178$ & $188$ & $1.90$ & $0.41$ \\
$(1,1)$ & $179$ & $182$ & $1.80$ & $0.34$ \\ 
$(2,0)$ & $179$ & $187$ & $1.94$ & $0.46$ \\ 
$(2,1)$ & $179$ & $185$ & $1.86$ & $0.41$ \\ 
$(2,2)$ & $180$ & $186$ & $1.79$ & $0.31$ \\
$(3,0)$ & $179$ & $187$ & $1.88$ & $0.37$ \\  
$(3,1)$ & $179$ & $187$ & $1.79$ & $0.32$ \\  
$(3,2)$ & $179$ & $187$ & $1.79$ & $0.31$ \\ 
$(3,3)$ & $180$ & $187$ & $1.79$ & $0.31$ \\
\end{tabular}
\label{tab2}
\end{ruledtabular}
\end{table}

The adsorption heights, Co magnetic moments and total energy differences between nonmagnetic and magnetic calculations are summarized in Table \ref{tab2}. 
The general trend in all systems under consideration is that the formation of magnetic moments on the order of $2\,\mu_\text{B}$ ($\mu_\text{B}$: Bohr magneton) is favorable by energies on the order of $\approx 0.5\,\text{eV}$ and associated with an upward relaxation of the Co atoms on the order of $10\,\text{pm}$\@. 

\begin{figure}
\includegraphics[width=0.8\columnwidth]{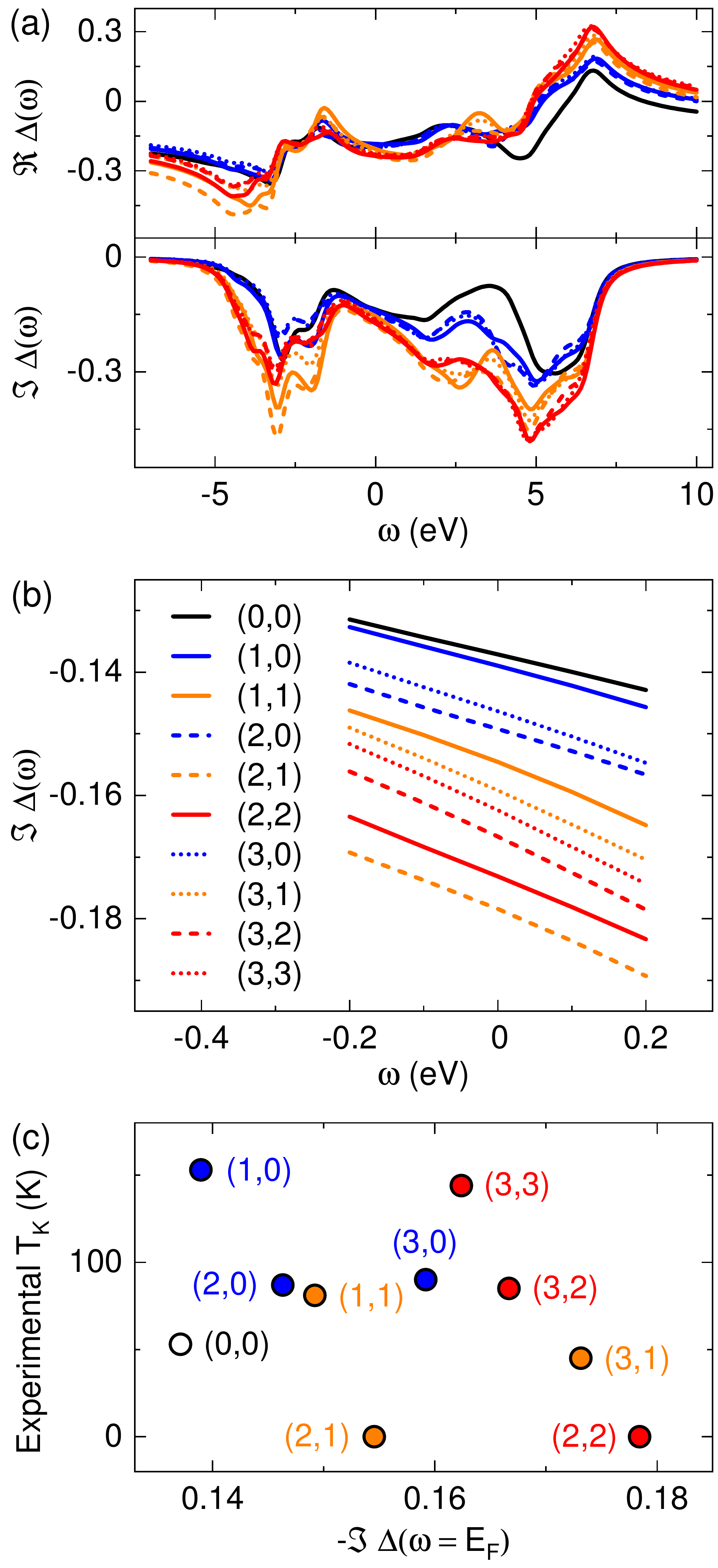}
\caption[fig5]{(Color online) 
Orbitally averaged hybridization functions $\Delta$ and correlation with $T_\text{K}$\@. 
(a) Real (top) and imaginary (bottom) parts of $\Delta(\omega)$ for the Co $d$ orbitals in the different $(m,n)$-clusters on Cu(111)\@. 
The legend is presented in (b)\@.
(b) Close-up view of $\Im\Delta(\omega)$ for energies around $E_\text{F}$\@. 
(c) Experimentally extracted $T_\text{K}$ \textit{versus} $-\Im\Delta(E_\text{F})$ where each data point corresponds to one $(m,n)$-cluster.
}
\label{fig5}
\end{figure}

A direct comparison of total energies without and with spin polarization as a function of adsorption height of the Co atoms in $(m,n)$-clusters is shown in Fig.\,\ref{fig4}(b)\@. 
We find that all clusters favor local magnetic moment formation around the Co sites regardless of structural relaxation details. 
Consequently, all $(m,n)$-clusters considered here are potential Kondo systems, including $(1,2)$, $(2,1)$, $(2,2)$-clusters for which an ASK resonance is not discernible in the experiments [Fig.\,\ref{fig2}(b)]\@.

\textit{Electronic structure in GGA and effective Anderson model.}
To investigate possible Kondo physics in the structures at hand, multi-orbital Anderson impurity models (AIM) were derived describing the $(m,n)$-clusters from the DFT-GGA calculations, where electronic correlations arise in the five Co $d$ orbitals, \textit{i.\,e.},
\begin{equation}
\label{eqn:H_AIM}
H_{\text{AIM}} = \sum_{k}\varepsilon_k c^\dagger_{k}c_{k}+\sum_{k,\alpha}\left(V_{k\alpha} c^\dagger_{k}d_{\alpha} + \text{H.c.}\right) + H_{\text{loc}}
\end{equation}
with
\begin{equation}
\label{eqn:H_LOC_AIM}
 H_{\text{loc}} = \sum_\sigma\varepsilon_{\alpha}d^\dagger_{\alpha}d_{\alpha}+ \frac{1}{2}\sum_{\alpha_1,...,\alpha_4} U_{\alpha_1,...,\alpha_4} d_{\alpha_1}^\dagger d_{\alpha_2}^\dagger d_{\alpha_3} d_{\alpha_4}.
\end{equation}
Here, the operators $d_{\alpha}$ ($d_{\alpha}^\dagger$) annihilate (create) electrons in Co $d$ orbitals characterized by orbital and spin quantum numbers $\alpha$, with corresponding on-site energies $\varepsilon_{\alpha}$, and local Coulomb interactions $U_{\alpha_1,...,\alpha_4}$. 
This impurity is embedded in a sea of conduction electron states originating from the surrounding Cu atoms and is described by annihilation (creation) operators $c_{k}$ ($c_{k}^\dagger$) and dispersions $\varepsilon_k$, where $k$ includes crystal momentum, band index and spin. 
The coupling between the impurity and the conduction electrons is provided by the hybridization $V_{k\alpha}$.

In the AIM only the impurity site is subject to an interaction term, while the bath of conduction electrons is assumed to be noninteracting.  
The bath degrees of freedom can thus be integrated out and the local electronic properties of the impurity can be described by $H_{\rm loc}$ and the hybridization function
\begin{equation}
\label{eqn:def_hybridization_function}
\Delta_{\alpha_1\alpha_2}(\omega)=\sum_{k}\frac{V_{k\alpha_1}^*V_{k\alpha_2}}{\omega + \text{i}\Gamma-\varepsilon_k}.
\end{equation}
We calculated $\Delta_{\alpha_1\alpha_2}(\omega)$ from first-principles following Refs.\,\onlinecite{prb_77_205112,jpcm_23_085601} with effective broadening $\Gamma=0.31\,\text{eV}$\@.
The orbital and spin diagonal elements of the hybridization function will enter the QMC simulations presented in the section below. 
For a first qualitative overview of the hybridization strength, which is one of the determining factors for the occurrence of the Kondo effect and associated to $T_\text{K}$, we show the orbitally averaged hybridization functions $\Delta(\omega)=\frac{1}{10}\sum_\alpha \Delta_{\alpha\alpha}(\omega)$ for all systems under consideration in Fig.\,\ref{fig5}(a),(b)\@.

While the hybridization functions differ between the individual chain realizations in several details, a few overarching statements can be made. 
First and foremost, the magnitude of the hybridization function and also the overall shape is similar in all cases and in line with the hybridization function found for Co atoms on Cu(111) from Ref.\,\onlinecite{prb_85_085114}. 
On a very rough qualitative level we would thus expect similar $T_\text{K}$ across the whole range of $(m,n)$-clusters.

Regarding variations between the different chain structures, we find an increase in hybridization with coordination number of the Co atoms, as one expects. 
A scaling analysis \cite{jp_41_193,cup_1993} suggests that $T_\text{K}$ depends on hybridization $\Delta=\Delta(E_\text{F})$ and local interaction $U$ according to 
\begin{equation}
T_\text{K}\propto\exp\left[-\frac{\pi U}{\gamma|\Im\,\Delta(E_\text{F}+\text{i}0^+)|}\right],
\label{eq:TKU}
\end{equation}
where $\gamma$ is a factor depending on orbital degeneracies. 
Therefore, increasing $|\Im\,\Delta(E_\text{F}+\text{i}0^+)|$ should result in (strongly) increasing $T_\text{K}$\@. 
Figure \ref{fig5}(c) shows the measured $T_\text{K}$ in relation to the calculated average hybridization function $|\Im\,\Delta(E_\text{F}+\text{i}0^+)|$\@.
 There is no clear correlation between $|\Im\,\Delta(E_\text{F}+\text{i}0^+)|$ and $T_\text{K}$\@. 
On the contrary, the $(2,2)$-cluster, which displays no signature of the Kondo effect in the experiments has the largest $|\Im\,\Delta(E_\text{F}+\text{i}0^+)|$ among all systems of this study. 
We thus conclude that mechanisms beyond this simple orbitally independent model should be at work in determining the Kondo physics at least in these two special cases.

In order to account for the complex interplay of charge, orbital, and spin fluctuations we now turn to QMC simulations of the full AIM with \textit{ab initio} derived hybridization functions.

\subsubsection{Quantum Monte Carlo Simulations}

Using the implementation of continuous-time QMC in hybridization expansion (CT-HYB) \cite{prl_97_076405,prb_74_155107} contained in the package w2dynamics, \cite{cpc_235_388} we solve the AIM given in Eqs.\,\eqref{eqn:H_AIM},\eqref{eqn:H_LOC_AIM} numerically exactly. 
The energy levels of the bath, its hybridization with the impurity,
encoded in the hybridization function of
Eq.\,\eqref{eqn:def_hybridization_function}, and the energy
differences between the local levels are taken from the DFT-GGA calculations presented in the previous section. 
We neglect the off-diagonal elements of the hybridization as they are of small magnitude. 
The fully spherically symmetric local Coulomb interaction tensor
$U_{\alpha_1,...,\alpha_4}$ was generated from interaction parameters
determined using the constrained random phase approximation. \cite{prb_83_121101,prl_109_146401}
The chemical potential was chosen to reach an occupation of $8$ electrons in the Co $d$ shell.

We use various indicators for the occurrence of the Kondo effect.
One is the behavior of the imaginary part of the self-energy at low frequencies (the so called ``first Matsubara rule'')\@. \cite{prb_86_155136}
This, together with the calculated quasiparticle weights and scattering rates, does not seem to allow a clear distinction of the chains exhibiting the Kondo effect experimentally from the others.

\begin{figure}
\includegraphics[width=0.8\linewidth]{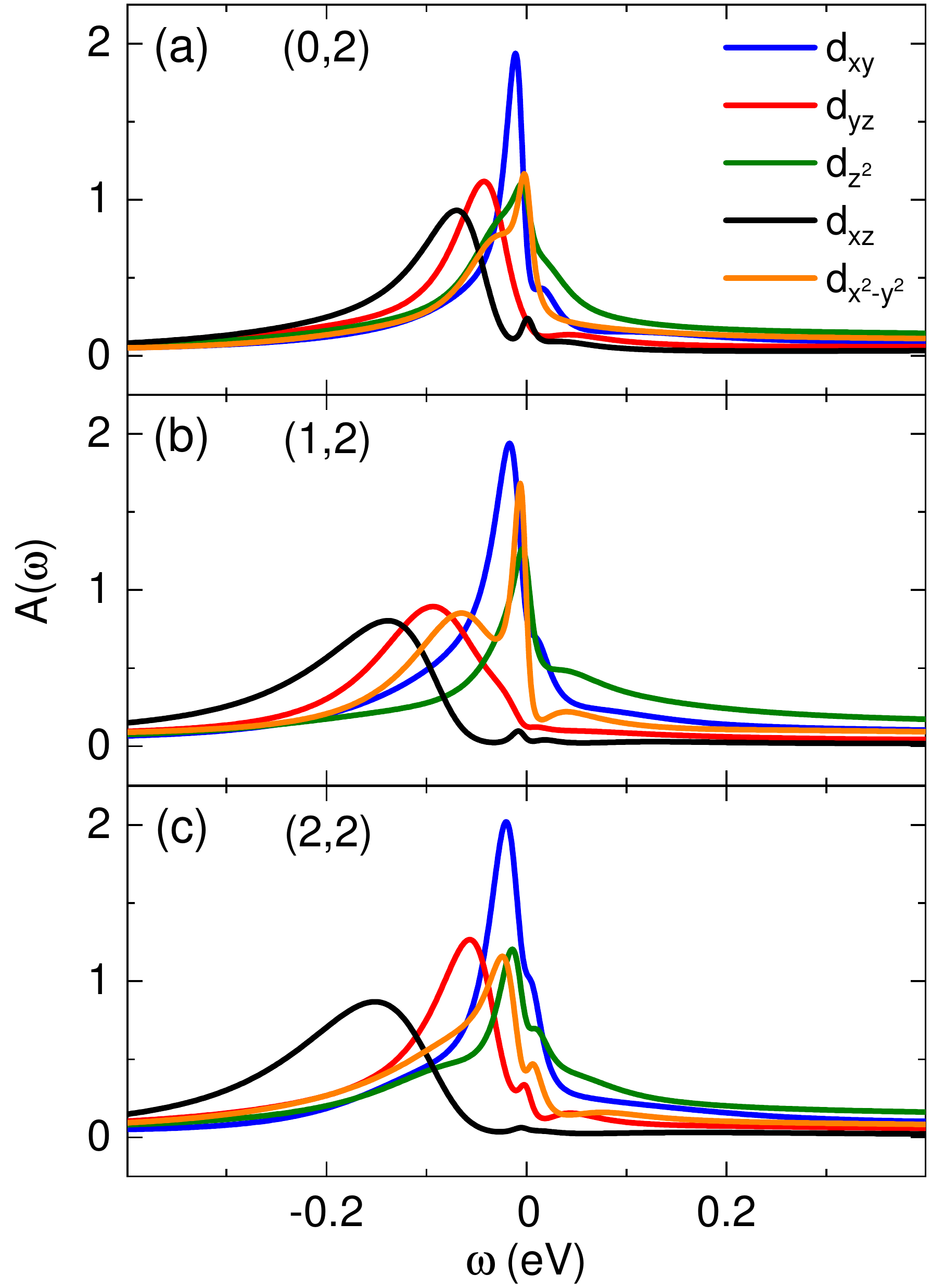}
\caption[fig6]{(Color online)
Comparison of orbitally resolved spectral functions.
Spectral functions $A(\omega)$ obtained for Co $d$ orbitals using the maximum entropy method from QMC simulations performed for $(m,n)$-clusters with (a) $m=0$, $n=2$; (b) $m=1$, $n=2$; (c)$m=2=n$ at a temperature of $46\,\text{K}$\@. 
}
\label{fig6}
\end{figure}

Another way of judging the propensity of exhibiting the Kondo effect is the direct inspection of the spectral function, $A(\omega)=-\frac{1}{\pi}\Im{G_\text{R}(\omega)}$\@.
Owing to the CT-HYB algorithm, $A(\omega)$ includes all local electronic correlation effects beyond DFT\@.
The full Green's function on the imaginary (discrete) Matsubara frequency axis obtained from CT-HYB was transformed to the retarded Green's function $G_\text{R}(\omega)$ on the real (continuous) frequency axis using the maximum-entropy method. \cite{pr_269_133,ebj_18_165,cpc_215_149}
According to Fig.\,\ref{fig6}, even considering many-body effects, chains that in the experiment have qualitatively different Kondo effects, behave quite similarly in theory. 
In particular, the $(0,2)$, $(1,2)$, $(2,2)$-clusters display a very similar orbital structure of the Kondo peaks, in spite of the fact that experimentally the first one is the only Kondo-active chain.

\begin{figure}
\includegraphics[width=0.9\linewidth]{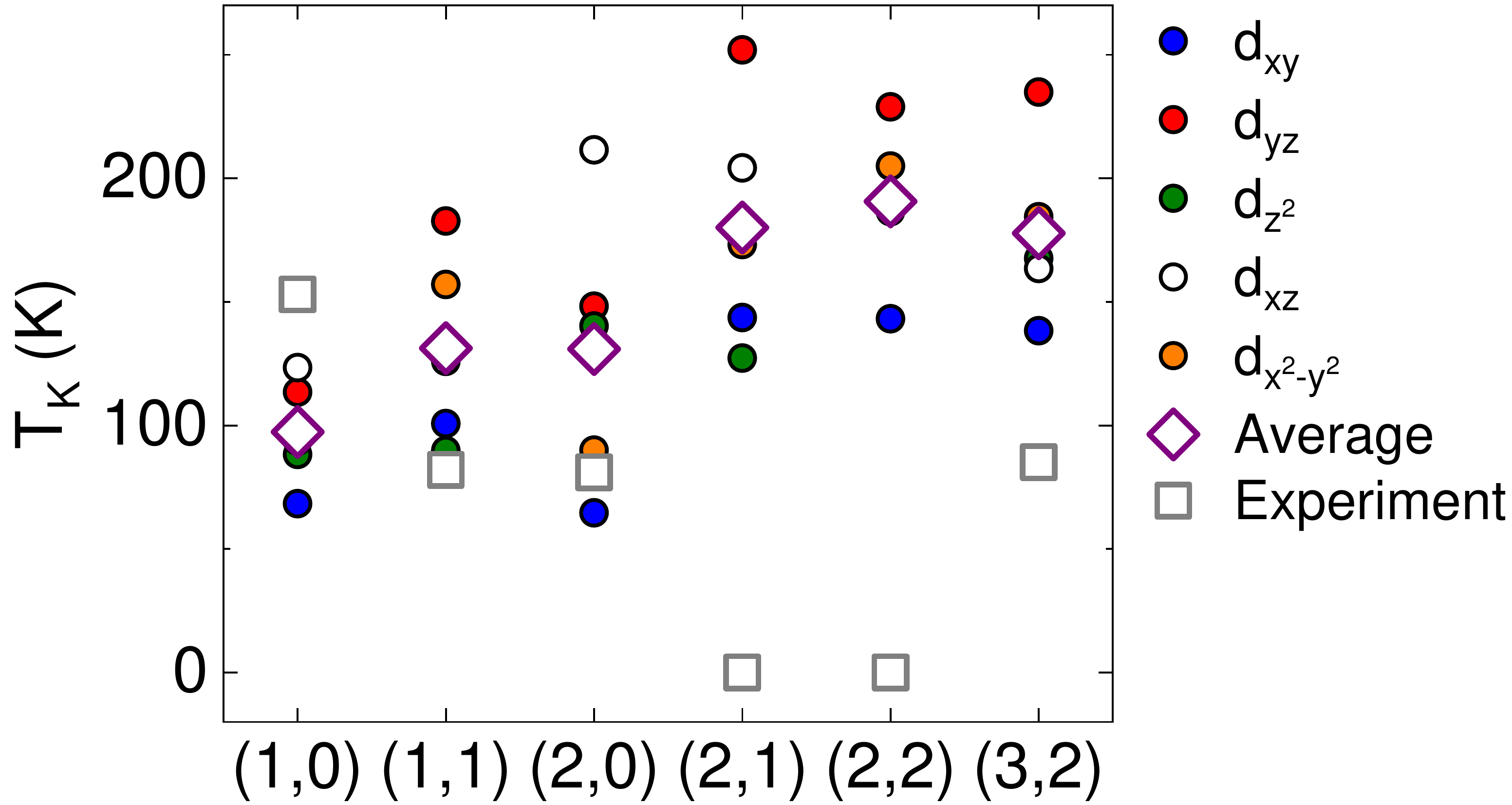}
\caption[fig7]{(Color online) 
Estimated orbitally resolved (dots) and averaged (lozenges) Kondo temperature $T_{\text{K}}$ for several $(m,n)$-clusters and comparison with experimental data (squares)\@.
The quasiparticle weight (see text) was calculated in the QMC simulations for each Co $d$ orbital and averaged.
The uncertainty margin of the calculated average spans the orbital variation, against which the statistical uncertainty is small.}
\label{fig7}
\end{figure}

To obtain a quantitative estimate of the Kondo temperature we use the
definition \cite{prb_85_085114,jpsj_74_8,jpcm_23_045601}
\begin{equation}
  \label{eq:TK_Z}
  T_{\text{K},\alpha} = -\frac{\pi}{4} Z \Im{\Delta_\alpha(0)}.
\end{equation}
The quasiparticle weight $Z$ was calculated from the imaginary part of $\Sigma(\mathrm{i}\omega_j)$, the Matsubara self-energy, as $Z={\left[1-\partial_{\omega_j}\Im{\Sigma(\mathrm{i}\omega_j)}\right]}^{-1}$ in the limit of vanishing $\mathrm{i}\omega_j$ ($j$: index of discrete frequency)\@.
In this way, $T_\text{K}$ may be calculated for each Co $d$ orbital in the $(m,n)$-clusters for which QMC simulations were performed.
Figure \ref{fig7} demonstrates that except for the cases in which no ASK resonance was observed experimentally, the calculated orbital-averaged $T_\text{K}$ [Eq.\,(\ref{eq:TK_Z})] never deviates from experimental values by more than a factor $2$\@.
Below we show that the orbital-dependent $T_{\text{K},\alpha}$ reacts sensitively to changes of some model parameters, which may be the origin of discrepancies between theory and experiment.

\begin{figure}
\includegraphics[width=0.8\linewidth]{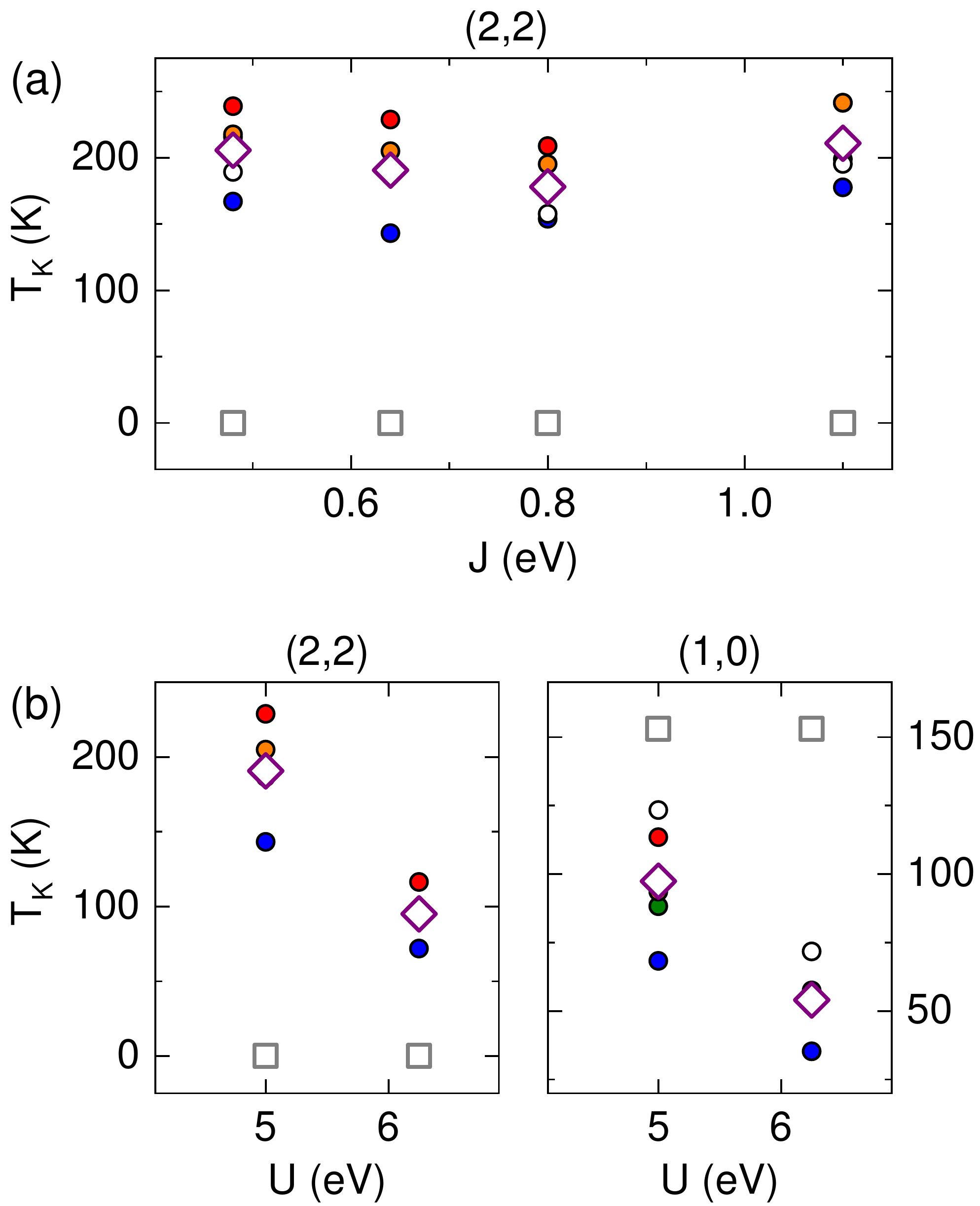}
\caption[fig8]{(Color online) 
Estimates of orbitally resolved (dots) and averaged (lozenges) $T_\text{K}$ of $(2,2)$, $(1,0)$-clusters from QMC quasiparticle-weight calculations and comparison with experimental data (squares)\@.
(a) Dependence of $T_\text{K}$ of a $(2,2)$-cluster on the Hund coupling constant $J$ for a Hubbard $U=5\,\text{eV}$\@.
(b) Dependence of $T_\text{K}$ of a $(2,2)$-cluster (left) and a $(1,0)$-cluster (right) on $U$ for $J=0.64\,\text{eV}$\@.
Uncertainty margins of the calculated average $T_\text{K}$ reflect the $T_\text{K}$ variations of the individual Co $d$ orbitals.
The dependence on $J$ (a) is nearly negligible while the dependence on $U$ (b) has roughly the same effect on the orbitally averaged value of both systems.
The legend is the same as in Fig.\,\ref{fig7}\@.
}
\label{fig8}
\end{figure}

\begin{figure}
\includegraphics[width=0.8\linewidth]{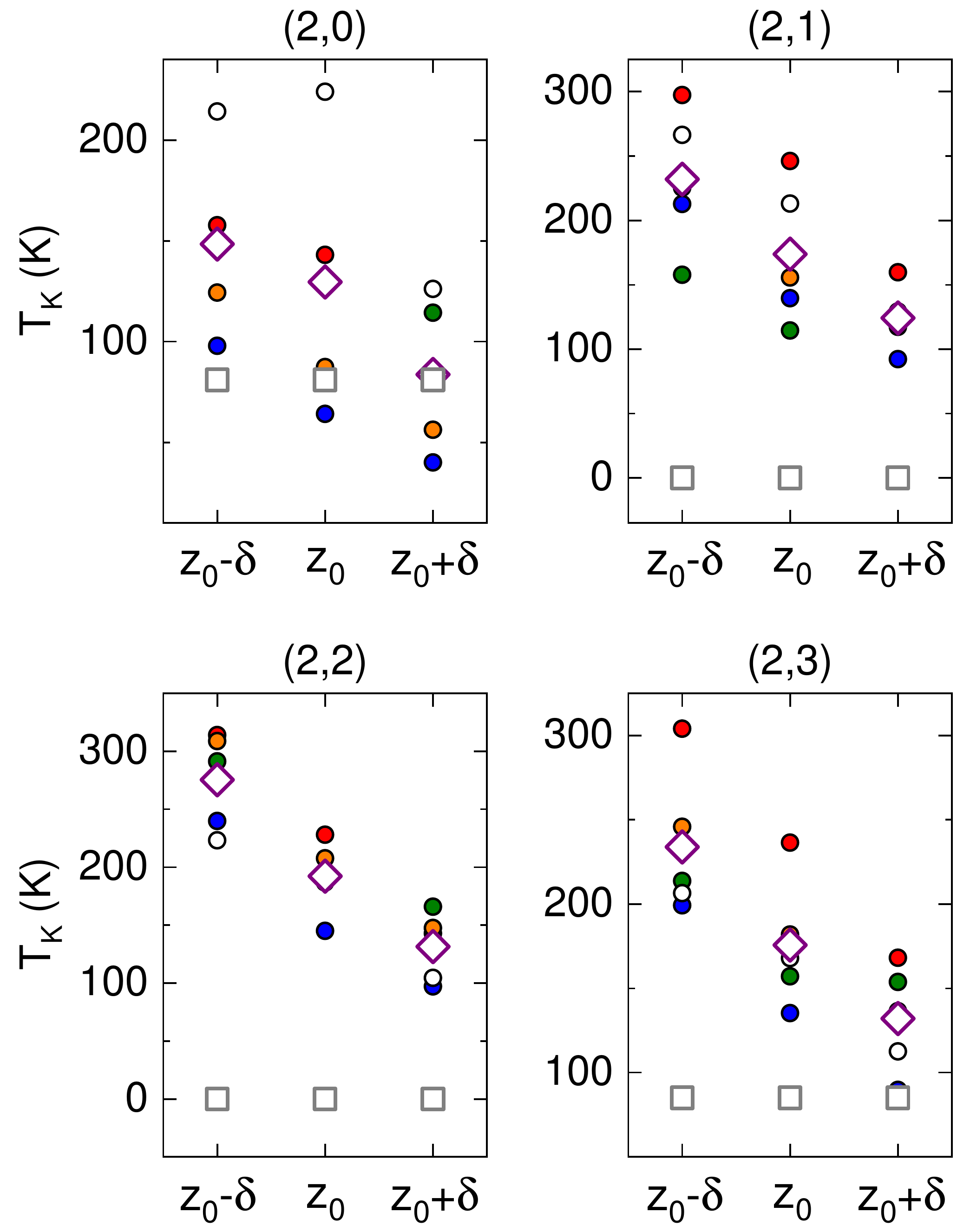}
\caption[fig9]{(Color online) 
Estimates of orbitally resolved (dots) and averaged (lozenges) $T_\text{K}$ of $(2,n)$-clusters $(0\leq n\leq 3)$ from QMC quasiparticle-weight calculations and comparison with experimental data (squares)\@.
The plots depict the dependence of $T_\text{K}$ on variations of the relaxed Co adsorption height $z_0$ by $\pm\delta$.
The uncertainty margins of the average $T_\text{K}$ reflect the $T_\text{K}$ variations of the individual Co $d$ orbitals, against which the statistical uncertainty is small. 
The dependence of the orbitally averaged value on the Co adsorption height is nearly the same for all considered clusters.
The legend is the same as in Fig.\,\ref{fig7}\@.}
\label{fig9}
\end{figure}

Experimentally relevant parameters of the AIM -- the intraorbital Hubbard interaction $U$ and the interorbital Hund coupling $J$ -- were varied in order to explore the sensitivity of $T_{\text{K},\alpha}$ to these parameters.
The representative example of a $(2,2)$-cluster shows that $T_{\text{K},\alpha}$ responds weakly to variations of $J$ [Fig.\,\ref{fig8}(a)]\@.
Increasing $J$ from $0.48\,\text{eV}$ to $1.10\,\text{eV}$ using $U=5\,\text{eV}$, the average of $T_{\text{K},\alpha}$ over all Co $d$ orbitals varies slightly between $170\,\text{K}$ and $211\,\text{K}$\@.
The response of $T_{\text{K},\alpha}$ to changes in $U$ is somewhat stronger as shown in Fig.\,\ref{fig8}(b) for $(2,2)$-clusters (left) and $(1,0)$-clusters (right)\@.
Using $J=0.64\,\text{eV}$ and increasing $U$ from $5\,\text{eV}$ to $6.25\,\text{eV}$ leads to a decrease of the averaged $T_{\text{K},\alpha}$ from $170\,\text{K}$ to $95\,\text{K}$ for $(2,2)$-clusters and from $97\,\text{K}$ to $54\,\text{K}$ for $(1,0)$-clusters, \text{i.\,e.}, by roughly a factor $2$ in both cases.

A final test was carried out to explore the dependence of $T_{\text{K},\alpha}$ on variations of the Co adsorption height by $\pm 10\,\text{pm}$.
The response of the entire spectral function and, in particular, of the ASK resonance energy to vertical displacements of the Co atom in the chains, is quantitatively different for ($m,n)$-clusters.
Figure \ref{fig9} shows that while quantitative changes in $T_{\text{K},\alpha}$ of individual orbitals are not the same, the averaged values exhibit very similar behavior for the different clusters.
Moreover, a simple relation to the experimental $T_\text{K}$ is not easy to recognize.

\section{Conclusion}  

Scanning tunneling spectroscopy of atom-by-atom assembled linear Cu$_m$CoCu$_n$ $(0\leq m,n\leq 5)$ clusters on Cu(111) reveals rich Kondo physics of the Co atom.
Describing the ASK resonance with a Fano line shape unravels a wide range of $T_\text{K}$ from $\approx 40\,\text{K}$ to $\approx 160\,\text{K}$ depending on $m$ and $n$.
Most strikingly, for CuCoCu$_2$ and Cu$_2$CoCu$_2$ the ASK resonance is entirely quenched hinting at the collapse of the Kondo effect.
 
DFT and QMC calculations indicate the presence of ASK resonances in the spectral functions of all linear clusters. 
From calculations for the different geometries estimates of $T_\text{K}$ range between $\approx 100\,\text{K}$ and $\approx 200\,\text{K}$\@. 
Under moderate variations of the interaction parameters or the height of the Co atom, the range of estimates extends further in both directions. 
The simulations thus capture $T_\text{K}$ qualitatively correctly for the majority of clusters. 
There are, however, clear deviations of the theoretical from the experimental results. 
First, the variations of $T_\text{K}$ between the different clusters is not correctly accounted for by the current level of theory. 
The most remarkable deviation is the presence of an ASK resonance for $(1,2)$, $(2,2)$-clusters in the simulations, clearly contradicting the experimental findings.
 
In order to identify the origin of these deviations several experimentally relevant model parameters, \textit{i.\,e.}, the electron--electron coupling strength in the Coulomb tensor, the Co orbital occupation and the Co atom adsorption height, were varied.
The qualitative response of the orbitally averaged $T_\text{K}$ is similar for all clusters suggesting that the experimentally observed absence of the ASK resonance for $(1,2)$, $(2,2)$-clusters is neither of simple electronic nor structral origin.
Rather, the microscopic explanation of the quenched ASK resonance in $(1,2)$, $(2,2)$-clusters may involve the increased size of the supercell for simulations, the transmission function to the tip, electron-phonon coupling, and an enlarged basis set for the description of low-energy, spin-orbit interaction and other relativistic effects.

From a methodological modeling standpoint, our results demonstrate that the complexity encoded in nanosystems like adsorbed atoms or clusters on surfaces combined with electronic correlation effects is substantial.
It presents a partially solved problem as regards quantitative aspects such as the occurrence of Kondo physics in most systems studied here and a generally qualitatively correct match between experimental and theoretical $T_\text{K}$\@. 
Yet, there are clear open issues -- in particular the variation of $T_\text{K}$ between the different systems. 
In our view, the investigated linear Cu$_m$CoCu$_n$ ($0\leq m, n\leq 5$) clusters on Cu(111) can serve as a very informative benchmark case for the future development of theoretical approaches to correlated nanosystems.

Moreover, the approach presented here may become important for describing Kondo screening in superconductors, which is relevant to topological quantum computation where it is crucial to manipulate and control chains of magnetically active atoms on superconductors. 
Rather than directly pursuing a solution of localized magnetic moments in a bath with Bardeen-Cooper-Schrieffer pairing fields, an alternative, as demonstrated in this work, is to work in a first-principle fully many-body framework including most of the realistic correlation effects and understand the problem \emph{before} going to a superconducting bath. 

\acknowledgments
Discussion with H.\ Kroha (Bonn) and sharing experimental data with R.\ Berndt (Kiel) prior to publication is acknowledged.
The work was supported by the Deutsche Forschungsgemeinschaft (DFG, German Research Foundation, Project-ID 258499086, SFB 1170), through the Würzburg-Dresden Cluster of Excellence on Complexity and Topology in Quantum Matter – ct.qmat (Project-ID 39085490, EXC 2147), and through the research training group Quantum Mechanical Materials Modelling (QM$^3$) (GRK 2247)\@. 
The authors acknowledge the Gauss Centre for Supercomputing e.\,V.\,for funding this project by providing computing time on the GCS Supercomputer SuperMUC at Leibniz Supercomputing Centre. 
Computing time provided by the North-German Supercomputing Alliance (HLRN) is acknowledged.

\bibliographystyle{apsrev4-1}

\begin{thebibliography}{100}%
\makeatletter
\providecommand \@ifxundefined [1]{%
 \@ifx{#1\undefined}
}%
\providecommand \@ifnum [1]{%
 \ifnum #1\expandafter \@firstoftwo
 \else \expandafter \@secondoftwo
 \fi
}%
\providecommand \@ifx [1]{%
 \ifx #1\expandafter \@firstoftwo
 \else \expandafter \@secondoftwo
 \fi
}%
\providecommand \natexlab [1]{#1}%
\providecommand \enquote  [1]{``#1''}%
\providecommand \bibnamefont  [1]{#1}%
\providecommand \bibfnamefont [1]{#1}%
\providecommand \citenamefont [1]{#1}%
\providecommand \href@noop [0]{\@secondoftwo}%
\providecommand \href [0]{\begingroup \@sanitize@url \@href}%
\providecommand \@href[1]{\@@startlink{#1}\@@href}%
\providecommand \@@href[1]{\endgroup#1\@@endlink}%
\providecommand \@sanitize@url [0]{\catcode `\\12\catcode `\$12\catcode
  `\&12\catcode `\#12\catcode `\^12\catcode `\_12\catcode `\%12\relax}%
\providecommand \@@startlink[1]{}%
\providecommand \@@endlink[0]{}%
\providecommand \url  [0]{\begingroup\@sanitize@url \@url }%
\providecommand \@url [1]{\endgroup\@href {#1}{\urlprefix }}%
\providecommand \urlprefix  [0]{URL }%
\providecommand \Eprint [0]{\href }%
\providecommand \doibase [0]{http://dx.doi.org/}%
\providecommand \selectlanguage [0]{\@gobble}%
\providecommand \bibinfo  [0]{\@secondoftwo}%
\providecommand \bibfield  [0]{\@secondoftwo}%
\providecommand \translation [1]{[#1]}%
\providecommand \BibitemOpen [0]{}%
\providecommand \bibitemStop [0]{}%
\providecommand \bibitemNoStop [0]{.\EOS\space}%
\providecommand \EOS [0]{\spacefactor3000\relax}%
\providecommand \BibitemShut  [1]{\csname bibitem#1\endcsname}%
\let\auto@bib@innerbib\@empty
\bibitem [{\citenamefont {Meissner}\ and\ \citenamefont
  {Voigt}(1930)}]{annphys_7_761}%
  \BibitemOpen
  \bibfield  {author} {\bibinfo {author} {\bibfnamefont {W.}~\bibnamefont
  {Meissner}}\ and\ \bibinfo {author} {\bibfnamefont {B.}~\bibnamefont
  {Voigt}},\ }\href {\doibase 10.1002/andp.19303990702} {\bibfield  {journal}
  {\bibinfo  {journal} {Annalen der Physik}\ }\textbf {\bibinfo {volume}
  {399}},\ \bibinfo {pages} {761} (\bibinfo {year} {1930})}\BibitemShut
  {NoStop}%
\bibitem [{\citenamefont {de~Haas}\ \emph {et~al.}(1934)\citenamefont
  {de~Haas}, \citenamefont {de~Boer},\ and\ \citenamefont {van~dën
  Berg}}]{physica_34_1115}%
  \BibitemOpen
  \bibfield  {author} {\bibinfo {author} {\bibfnamefont {W.}~\bibnamefont
  {de~Haas}}, \bibinfo {author} {\bibfnamefont {J.}~\bibnamefont {de~Boer}}, \
  and\ \bibinfo {author} {\bibfnamefont {G.}~\bibnamefont {van~dën Berg}},\
  }\href {\doibase https://doi.org/10.1016/S0031-8914(34)80310-2} {\bibfield
  {journal} {\bibinfo  {journal} {Physica}\ }\textbf {\bibinfo {volume} {1}},\
  \bibinfo {pages} {1115 } (\bibinfo {year} {1934})}\BibitemShut {NoStop}%
\bibitem [{\citenamefont {Kondo}(1964)}]{ptp_32_37}%
  \BibitemOpen
  \bibfield  {author} {\bibinfo {author} {\bibfnamefont {J.}~\bibnamefont
  {Kondo}},\ }\href {\doibase 10.1143/PTP.32.37} {\bibfield  {journal}
  {\bibinfo  {journal} {Progress of Theoretical Physics}\ }\textbf {\bibinfo
  {volume} {32}},\ \bibinfo {pages} {37} (\bibinfo {year} {1964})}\BibitemShut
  {NoStop}%
\bibitem [{\citenamefont {Kondo}(1968)}]{pr_169_437}%
  \BibitemOpen
  \bibfield  {author} {\bibinfo {author} {\bibfnamefont {J.}~\bibnamefont
  {Kondo}},\ }\href {\doibase 10.1103/PhysRev.169.437} {\bibfield  {journal}
  {\bibinfo  {journal} {Phys. Rev.}\ }\textbf {\bibinfo {volume} {169}},\
  \bibinfo {pages} {437} (\bibinfo {year} {1968})}\BibitemShut {NoStop}%
\bibitem [{\citenamefont {Anderson}(1961)}]{pr_124_41}%
  \BibitemOpen
  \bibfield  {author} {\bibinfo {author} {\bibfnamefont {P.~W.}\ \bibnamefont
  {Anderson}},\ }\href {\doibase 10.1103/PhysRev.124.41} {\bibfield  {journal}
  {\bibinfo  {journal} {Phys. Rev.}\ }\textbf {\bibinfo {volume} {124}},\
  \bibinfo {pages} {41} (\bibinfo {year} {1961})}\BibitemShut {NoStop}%
\bibitem [{\citenamefont {Anderson}(1970)}]{jpcssp_3_2436}%
  \BibitemOpen
  \bibfield  {author} {\bibinfo {author} {\bibfnamefont {P.~W.}\ \bibnamefont
  {Anderson}},\ }\href {\doibase 10.1088/0022-3719/3/12/008} {\bibfield
  {journal} {\bibinfo  {journal} {Journal of Physics C: Solid State Physics}\
  }\textbf {\bibinfo {volume} {3}},\ \bibinfo {pages} {2436} (\bibinfo {year}
  {1970})}\BibitemShut {NoStop}%
\bibitem [{\citenamefont {Wilson}(1975)}]{rmp_47_773}%
  \BibitemOpen
  \bibfield  {author} {\bibinfo {author} {\bibfnamefont {K.~G.}\ \bibnamefont
  {Wilson}},\ }\href {\doibase 10.1103/RevModPhys.47.773} {\bibfield  {journal}
  {\bibinfo  {journal} {Rev. Mod. Phys.}\ }\textbf {\bibinfo {volume} {47}},\
  \bibinfo {pages} {773} (\bibinfo {year} {1975})}\BibitemShut {NoStop}%
\bibitem [{\citenamefont {Abrikosov}(1965{\natexlab{a}})}]{ppf_2_5}%
  \BibitemOpen
  \bibfield  {author} {\bibinfo {author} {\bibfnamefont {A.~A.}\ \bibnamefont
  {Abrikosov}},\ }\href {\doibase 10.1103/PhysicsPhysiqueFizika.2.5} {\bibfield
   {journal} {\bibinfo  {journal} {Physics Physique Fizika}\ }\textbf {\bibinfo
  {volume} {2}},\ \bibinfo {pages} {5} (\bibinfo {year}
  {1965}{\natexlab{a}})}\BibitemShut {NoStop}%
\bibitem [{\citenamefont {Abrikosov}(1965{\natexlab{b}})}]{ppf_2_61}%
  \BibitemOpen
  \bibfield  {author} {\bibinfo {author} {\bibfnamefont {A.~A.}\ \bibnamefont
  {Abrikosov}},\ }\href {\doibase 10.1103/PhysicsPhysiqueFizika.2.61}
  {\bibfield  {journal} {\bibinfo  {journal} {Physics Physique Fizika}\
  }\textbf {\bibinfo {volume} {2}},\ \bibinfo {pages} {61} (\bibinfo {year}
  {1965}{\natexlab{b}})}\BibitemShut {NoStop}%
\bibitem [{\citenamefont {Suhl}(1967)}]{aca_1967}%
  \BibitemOpen
  \bibfield  {author} {\bibinfo {author} {\bibfnamefont {H.}~\bibnamefont
  {Suhl}},\ }\enquote {\bibinfo {title} {Theory of magnetism in transition
  metals},}\ \ (\bibinfo  {publisher} {Academic, London},\ \bibinfo {year}
  {1967})\ pp.\ \bibinfo {pages} {116 -- 205}\BibitemShut {NoStop}%
\bibitem [{\citenamefont {Patthey}\ \emph {et~al.}(1985)\citenamefont
  {Patthey}, \citenamefont {Delley}, \citenamefont {Schneider},\ and\
  \citenamefont {Baer}}]{prl_55_1518}%
  \BibitemOpen
  \bibfield  {author} {\bibinfo {author} {\bibfnamefont {F.}~\bibnamefont
  {Patthey}}, \bibinfo {author} {\bibfnamefont {B.}~\bibnamefont {Delley}},
  \bibinfo {author} {\bibfnamefont {W.~D.}\ \bibnamefont {Schneider}}, \ and\
  \bibinfo {author} {\bibfnamefont {Y.}~\bibnamefont {Baer}},\ }\href {\doibase
  10.1103/PhysRevLett.55.1518} {\bibfield  {journal} {\bibinfo  {journal}
  {Phys. Rev. Lett.}\ }\textbf {\bibinfo {volume} {55}},\ \bibinfo {pages}
  {1518} (\bibinfo {year} {1985})}\BibitemShut {NoStop}%
\bibitem [{\citenamefont {Patthey}\ \emph {et~al.}(1987)\citenamefont
  {Patthey}, \citenamefont {Schneider}, \citenamefont {Baer},\ and\
  \citenamefont {Delley}}]{prl_58_2810}%
  \BibitemOpen
  \bibfield  {author} {\bibinfo {author} {\bibfnamefont {F.}~\bibnamefont
  {Patthey}}, \bibinfo {author} {\bibfnamefont {W.~D.}\ \bibnamefont
  {Schneider}}, \bibinfo {author} {\bibfnamefont {Y.}~\bibnamefont {Baer}}, \
  and\ \bibinfo {author} {\bibfnamefont {B.}~\bibnamefont {Delley}},\ }\href
  {\doibase 10.1103/PhysRevLett.58.2810} {\bibfield  {journal} {\bibinfo
  {journal} {Phys. Rev. Lett.}\ }\textbf {\bibinfo {volume} {58}},\ \bibinfo
  {pages} {2810} (\bibinfo {year} {1987})}\BibitemShut {NoStop}%
\bibitem [{\citenamefont {Patthey}\ \emph {et~al.}(1990)\citenamefont
  {Patthey}, \citenamefont {Imer}, \citenamefont {Schneider}, \citenamefont
  {Beck}, \citenamefont {Baer},\ and\ \citenamefont {Delley}}]{prb_42_8864}%
  \BibitemOpen
  \bibfield  {author} {\bibinfo {author} {\bibfnamefont {F.}~\bibnamefont
  {Patthey}}, \bibinfo {author} {\bibfnamefont {J.-M.}\ \bibnamefont {Imer}},
  \bibinfo {author} {\bibfnamefont {W.-D.}\ \bibnamefont {Schneider}}, \bibinfo
  {author} {\bibfnamefont {H.}~\bibnamefont {Beck}}, \bibinfo {author}
  {\bibfnamefont {Y.}~\bibnamefont {Baer}}, \ and\ \bibinfo {author}
  {\bibfnamefont {B.}~\bibnamefont {Delley}},\ }\href {\doibase
  10.1103/PhysRevB.42.8864} {\bibfield  {journal} {\bibinfo  {journal} {Phys.
  Rev. B}\ }\textbf {\bibinfo {volume} {42}},\ \bibinfo {pages} {8864}
  (\bibinfo {year} {1990})}\BibitemShut {NoStop}%
\bibitem [{\citenamefont {Laubschat}\ \emph {et~al.}(1990)\citenamefont
  {Laubschat}, \citenamefont {Weschke}, \citenamefont {Holtz}, \citenamefont
  {Domke}, \citenamefont {Strebel},\ and\ \citenamefont
  {Kaindl}}]{prl_65_1639}%
  \BibitemOpen
  \bibfield  {author} {\bibinfo {author} {\bibfnamefont {C.}~\bibnamefont
  {Laubschat}}, \bibinfo {author} {\bibfnamefont {E.}~\bibnamefont {Weschke}},
  \bibinfo {author} {\bibfnamefont {C.}~\bibnamefont {Holtz}}, \bibinfo
  {author} {\bibfnamefont {M.}~\bibnamefont {Domke}}, \bibinfo {author}
  {\bibfnamefont {O.}~\bibnamefont {Strebel}}, \ and\ \bibinfo {author}
  {\bibfnamefont {G.}~\bibnamefont {Kaindl}},\ }\href {\doibase
  10.1103/PhysRevLett.65.1639} {\bibfield  {journal} {\bibinfo  {journal}
  {Phys. Rev. Lett.}\ }\textbf {\bibinfo {volume} {65}},\ \bibinfo {pages}
  {1639} (\bibinfo {year} {1990})}\BibitemShut {NoStop}%
\bibitem [{\citenamefont {Weschke}\ \emph {et~al.}(1991)\citenamefont
  {Weschke}, \citenamefont {Laubschat}, \citenamefont {Simmons}, \citenamefont
  {Domke}, \citenamefont {Strebel},\ and\ \citenamefont
  {Kaindl}}]{prb_44_8304}%
  \BibitemOpen
  \bibfield  {author} {\bibinfo {author} {\bibfnamefont {E.}~\bibnamefont
  {Weschke}}, \bibinfo {author} {\bibfnamefont {C.}~\bibnamefont {Laubschat}},
  \bibinfo {author} {\bibfnamefont {T.}~\bibnamefont {Simmons}}, \bibinfo
  {author} {\bibfnamefont {M.}~\bibnamefont {Domke}}, \bibinfo {author}
  {\bibfnamefont {O.}~\bibnamefont {Strebel}}, \ and\ \bibinfo {author}
  {\bibfnamefont {G.}~\bibnamefont {Kaindl}},\ }\href {\doibase
  10.1103/PhysRevB.44.8304} {\bibfield  {journal} {\bibinfo  {journal} {Phys.
  Rev. B}\ }\textbf {\bibinfo {volume} {44}},\ \bibinfo {pages} {8304}
  (\bibinfo {year} {1991})}\BibitemShut {NoStop}%
\bibitem [{\citenamefont {Ehm}\ \emph {et~al.}(2007)\citenamefont {Ehm},
  \citenamefont {H\"ufner}, \citenamefont {Reinert}, \citenamefont {Kroha},
  \citenamefont {W\"olfle}, \citenamefont {Stockert}, \citenamefont {Geibel},\
  and\ \citenamefont {L\"ohneysen}}]{prb_76_045117}%
  \BibitemOpen
  \bibfield  {author} {\bibinfo {author} {\bibfnamefont {D.}~\bibnamefont
  {Ehm}}, \bibinfo {author} {\bibfnamefont {S.}~\bibnamefont {H\"ufner}},
  \bibinfo {author} {\bibfnamefont {F.}~\bibnamefont {Reinert}}, \bibinfo
  {author} {\bibfnamefont {J.}~\bibnamefont {Kroha}}, \bibinfo {author}
  {\bibfnamefont {P.}~\bibnamefont {W\"olfle}}, \bibinfo {author}
  {\bibfnamefont {O.}~\bibnamefont {Stockert}}, \bibinfo {author}
  {\bibfnamefont {C.}~\bibnamefont {Geibel}}, \ and\ \bibinfo {author}
  {\bibfnamefont {H.~v.}\ \bibnamefont {L\"ohneysen}},\ }\href {\doibase
  10.1103/PhysRevB.76.045117} {\bibfield  {journal} {\bibinfo  {journal} {Phys.
  Rev. B}\ }\textbf {\bibinfo {volume} {76}},\ \bibinfo {pages} {045117}
  (\bibinfo {year} {2007})}\BibitemShut {NoStop}%
\bibitem [{\citenamefont {Wuilloud}\ \emph {et~al.}(1983)\citenamefont
  {Wuilloud}, \citenamefont {Moser}, \citenamefont {Schneider},\ and\
  \citenamefont {Baer}}]{prb_28_7354}%
  \BibitemOpen
  \bibfield  {author} {\bibinfo {author} {\bibfnamefont {E.}~\bibnamefont
  {Wuilloud}}, \bibinfo {author} {\bibfnamefont {H.~R.}\ \bibnamefont {Moser}},
  \bibinfo {author} {\bibfnamefont {W.~D.}\ \bibnamefont {Schneider}}, \ and\
  \bibinfo {author} {\bibfnamefont {Y.}~\bibnamefont {Baer}},\ }\href {\doibase
  10.1103/PhysRevB.28.7354} {\bibfield  {journal} {\bibinfo  {journal} {Phys.
  Rev. B}\ }\textbf {\bibinfo {volume} {28}},\ \bibinfo {pages} {7354}
  (\bibinfo {year} {1983})}\BibitemShut {NoStop}%
\bibitem [{\citenamefont {Li}\ \emph {et~al.}(1998)\citenamefont {Li},
  \citenamefont {Schneider}, \citenamefont {Berndt},\ and\ \citenamefont
  {Delley}}]{prl_80_2893}%
  \BibitemOpen
  \bibfield  {author} {\bibinfo {author} {\bibfnamefont {J.}~\bibnamefont
  {Li}}, \bibinfo {author} {\bibfnamefont {W.-D.}\ \bibnamefont {Schneider}},
  \bibinfo {author} {\bibfnamefont {R.}~\bibnamefont {Berndt}}, \ and\ \bibinfo
  {author} {\bibfnamefont {B.}~\bibnamefont {Delley}},\ }\href {\doibase
  10.1103/PhysRevLett.80.2893} {\bibfield  {journal} {\bibinfo  {journal}
  {Phys. Rev. Lett.}\ }\textbf {\bibinfo {volume} {80}},\ \bibinfo {pages}
  {2893} (\bibinfo {year} {1998})}\BibitemShut {NoStop}%
\bibitem [{\citenamefont {Madhavan}\ \emph {et~al.}(1998)\citenamefont
  {Madhavan}, \citenamefont {Chen}, \citenamefont {Jamneala}, \citenamefont
  {Crommie},\ and\ \citenamefont {Wingreen}}]{science_280_567}%
  \BibitemOpen
  \bibfield  {author} {\bibinfo {author} {\bibfnamefont {V.}~\bibnamefont
  {Madhavan}}, \bibinfo {author} {\bibfnamefont {W.}~\bibnamefont {Chen}},
  \bibinfo {author} {\bibfnamefont {T.}~\bibnamefont {Jamneala}}, \bibinfo
  {author} {\bibfnamefont {M.~F.}\ \bibnamefont {Crommie}}, \ and\ \bibinfo
  {author} {\bibfnamefont {N.~S.}\ \bibnamefont {Wingreen}},\ }\href {\doibase
  10.1126/science.280.5363.567} {\bibfield  {journal} {\bibinfo  {journal}
  {Science}\ }\textbf {\bibinfo {volume} {280}},\ \bibinfo {pages} {567}
  (\bibinfo {year} {1998})}\BibitemShut {NoStop}%
\bibitem [{\citenamefont {Fano}(1935)}]{nc_12_154}%
  \BibitemOpen
  \bibfield  {author} {\bibinfo {author} {\bibfnamefont {U.}~\bibnamefont
  {Fano}},\ }\href {\doibase 10.1007/BF02958288} {\bibfield  {journal}
  {\bibinfo  {journal} {Il Nuovo Cimento}\ }\textbf {\bibinfo
  {volume} {12}},\ \bibinfo {pages} {154} (\bibinfo {year} {1935})}\BibitemShut
  {NoStop}%
\bibitem [{\citenamefont {Fano}(1961)}]{pr_124_1866}%
  \BibitemOpen
  \bibfield  {author} {\bibinfo {author} {\bibfnamefont {U.}~\bibnamefont
  {Fano}},\ }\href {\doibase 10.1103/PhysRev.124.1866} {\bibfield  {journal}
  {\bibinfo  {journal} {Phys. Rev.}\ }\textbf {\bibinfo {volume} {124}},\
  \bibinfo {pages} {1866} (\bibinfo {year} {1961})}\BibitemShut {NoStop}%
\bibitem [{\citenamefont {Frota}\ and\ \citenamefont
  {Oliveira}(1986)}]{prb_33_7871}%
  \BibitemOpen
  \bibfield  {author} {\bibinfo {author} {\bibfnamefont {H.~O.}\ \bibnamefont
  {Frota}}\ and\ \bibinfo {author} {\bibfnamefont {L.~N.}\ \bibnamefont
  {Oliveira}},\ }\href {\doibase 10.1103/PhysRevB.33.7871} {\bibfield
  {journal} {\bibinfo  {journal} {Phys. Rev. B}\ }\textbf {\bibinfo {volume}
  {33}},\ \bibinfo {pages} {7871} (\bibinfo {year} {1986})}\BibitemShut
  {NoStop}%
\bibitem [{\citenamefont {Frota}(1992)}]{prb_45_1096}%
  \BibitemOpen
  \bibfield  {author} {\bibinfo {author} {\bibfnamefont {H.~O.}\ \bibnamefont
  {Frota}},\ }\href {\doibase 10.1103/PhysRevB.45.1096} {\bibfield  {journal}
  {\bibinfo  {journal} {Phys. Rev. B}\ }\textbf {\bibinfo {volume} {45}},\
  \bibinfo {pages} {1096} (\bibinfo {year} {1992})}\BibitemShut {NoStop}%
\bibitem [{\citenamefont {\ifmmode~\check{Z}\else
  \v{Z}\fi{}itko}(2011)}]{prb_84_195116}%
  \BibitemOpen
  \bibfield  {author} {\bibinfo {author} {\bibfnamefont {R.}~\bibnamefont
  {\ifmmode~\check{Z}\else \v{Z}\fi{}itko}},\ }\href {\doibase
  10.1103/PhysRevB.84.195116} {\bibfield  {journal} {\bibinfo  {journal} {Phys.
  Rev. B}\ }\textbf {\bibinfo {volume} {84}},\ \bibinfo {pages} {195116}
  (\bibinfo {year} {2011})}\BibitemShut {NoStop}%
\bibitem [{\citenamefont {Pr\"user}\ \emph {et~al.}(2011)\citenamefont
  {Pr\"user}, \citenamefont {Wenderoth}, \citenamefont {Dargel}, \citenamefont
  {Weismann}, \citenamefont {Peters}, \citenamefont {Pruschke},\ and\
  \citenamefont {Ulbrich}}]{natphys_7_203}%
  \BibitemOpen
  \bibfield  {author} {\bibinfo {author} {\bibfnamefont {H.}~\bibnamefont
  {Pr\"user}}, \bibinfo {author} {\bibfnamefont {M.}~\bibnamefont {Wenderoth}},
  \bibinfo {author} {\bibfnamefont {P.~E.}\ \bibnamefont {Dargel}}, \bibinfo
  {author} {\bibfnamefont {A.}~\bibnamefont {Weismann}}, \bibinfo {author}
  {\bibfnamefont {R.}~\bibnamefont {Peters}}, \bibinfo {author} {\bibfnamefont
  {T.}~\bibnamefont {Pruschke}}, \ and\ \bibinfo {author} {\bibfnamefont
  {R.~G.}\ \bibnamefont {Ulbrich}},\ }\href@noop {} {\bibfield  {journal}
  {\bibinfo  {journal} {Nat. Phys.}\ }\textbf {\bibinfo {volume} {7}},\
  \bibinfo {pages} {203 } (\bibinfo {year} {2011})}\BibitemShut {NoStop}%
\bibitem [{\citenamefont {Pr\"user}\ \emph {et~al.}(2012)\citenamefont
  {Pr\"user}, \citenamefont {Wenderoth}, \citenamefont {Weismann},\ and\
  \citenamefont {Ulbrich}}]{prl_108_166604}%
  \BibitemOpen
  \bibfield  {author} {\bibinfo {author} {\bibfnamefont {H.}~\bibnamefont
  {Pr\"user}}, \bibinfo {author} {\bibfnamefont {M.}~\bibnamefont {Wenderoth}},
  \bibinfo {author} {\bibfnamefont {A.}~\bibnamefont {Weismann}}, \ and\
  \bibinfo {author} {\bibfnamefont {R.~G.}\ \bibnamefont {Ulbrich}},\ }\href
  {\doibase 10.1103/PhysRevLett.108.166604} {\bibfield  {journal} {\bibinfo
  {journal} {Phys. Rev. Lett.}\ }\textbf {\bibinfo {volume} {108}},\ \bibinfo
  {pages} {166604} (\bibinfo {year} {2012})}\BibitemShut {NoStop}%
\bibitem [{\citenamefont {Nagaoka}\ \emph {et~al.}(2002)\citenamefont
  {Nagaoka}, \citenamefont {Jamneala}, \citenamefont {Grobis},\ and\
  \citenamefont {Crommie}}]{prl_88_077205}%
  \BibitemOpen
  \bibfield  {author} {\bibinfo {author} {\bibfnamefont {K.}~\bibnamefont
  {Nagaoka}}, \bibinfo {author} {\bibfnamefont {T.}~\bibnamefont {Jamneala}},
  \bibinfo {author} {\bibfnamefont {M.}~\bibnamefont {Grobis}}, \ and\ \bibinfo
  {author} {\bibfnamefont {M.~F.}\ \bibnamefont {Crommie}},\ }\href {\doibase
  10.1103/PhysRevLett.88.077205} {\bibfield  {journal} {\bibinfo  {journal}
  {Phys. Rev. Lett.}\ }\textbf {\bibinfo {volume} {88}},\ \bibinfo {pages}
  {077205} (\bibinfo {year} {2002})}\BibitemShut {NoStop}%
\bibitem [{\citenamefont {Knorr}\ \emph {et~al.}(2002)\citenamefont {Knorr},
  \citenamefont {Schneider}, \citenamefont {Diekh\"oner}, \citenamefont
  {Wahl},\ and\ \citenamefont {Kern}}]{prl_88_096804}%
  \BibitemOpen
  \bibfield  {author} {\bibinfo {author} {\bibfnamefont {N.}~\bibnamefont
  {Knorr}}, \bibinfo {author} {\bibfnamefont {M.~A.}\ \bibnamefont
  {Schneider}}, \bibinfo {author} {\bibfnamefont {L.}~\bibnamefont
  {Diekh\"oner}}, \bibinfo {author} {\bibfnamefont {P.}~\bibnamefont {Wahl}}, \
  and\ \bibinfo {author} {\bibfnamefont {K.}~\bibnamefont {Kern}},\ }\href
  {\doibase 10.1103/PhysRevLett.88.096804} {\bibfield  {journal} {\bibinfo
  {journal} {Phys. Rev. Lett.}\ }\textbf {\bibinfo {volume} {88}},\ \bibinfo
  {pages} {096804} (\bibinfo {year} {2002})}\BibitemShut {NoStop}%
\bibitem [{\citenamefont {Wahl}\ \emph {et~al.}(2004)\citenamefont {Wahl},
  \citenamefont {Diekh\"oner}, \citenamefont {Schneider}, \citenamefont
  {Vitali}, \citenamefont {Wittich},\ and\ \citenamefont
  {Kern}}]{prl_93_176603}%
  \BibitemOpen
  \bibfield  {author} {\bibinfo {author} {\bibfnamefont {P.}~\bibnamefont
  {Wahl}}, \bibinfo {author} {\bibfnamefont {L.}~\bibnamefont {Diekh\"oner}},
  \bibinfo {author} {\bibfnamefont {M.~A.}\ \bibnamefont {Schneider}}, \bibinfo
  {author} {\bibfnamefont {L.}~\bibnamefont {Vitali}}, \bibinfo {author}
  {\bibfnamefont {G.}~\bibnamefont {Wittich}}, \ and\ \bibinfo {author}
  {\bibfnamefont {K.}~\bibnamefont {Kern}},\ }\href {\doibase
  10.1103/PhysRevLett.93.176603} {\bibfield  {journal} {\bibinfo  {journal}
  {Phys. Rev. Lett.}\ }\textbf {\bibinfo {volume} {93}},\ \bibinfo {pages}
  {176603} (\bibinfo {year} {2004})}\BibitemShut {NoStop}%
\bibitem [{\citenamefont {Choi}\ \emph {et~al.}(2012)\citenamefont {Choi},
  \citenamefont {Rastei}, \citenamefont {Simon},\ and\ \citenamefont
  {Limot}}]{prl_108_266803}%
  \BibitemOpen
  \bibfield  {author} {\bibinfo {author} {\bibfnamefont {D.-J.}\ \bibnamefont
  {Choi}}, \bibinfo {author} {\bibfnamefont {M.~V.}\ \bibnamefont {Rastei}},
  \bibinfo {author} {\bibfnamefont {P.}~\bibnamefont {Simon}}, \ and\ \bibinfo
  {author} {\bibfnamefont {L.}~\bibnamefont {Limot}},\ }\href {\doibase
  10.1103/PhysRevLett.108.266803} {\bibfield  {journal} {\bibinfo  {journal}
  {Phys. Rev. Lett.}\ }\textbf {\bibinfo {volume} {108}},\ \bibinfo {pages}
  {266803} (\bibinfo {year} {2012})}\BibitemShut {NoStop}%
\bibitem [{\citenamefont {von Bergmann}\ \emph {et~al.}(2015)\citenamefont {von
  Bergmann}, \citenamefont {Ternes}, \citenamefont {Loth}, \citenamefont
  {Lutz},\ and\ \citenamefont {Heinrich}}]{prl_114_076601}%
  \BibitemOpen
  \bibfield  {author} {\bibinfo {author} {\bibfnamefont {K.}~\bibnamefont {von
  Bergmann}}, \bibinfo {author} {\bibfnamefont {M.}~\bibnamefont {Ternes}},
  \bibinfo {author} {\bibfnamefont {S.}~\bibnamefont {Loth}}, \bibinfo {author}
  {\bibfnamefont {C.~P.}\ \bibnamefont {Lutz}}, \ and\ \bibinfo {author}
  {\bibfnamefont {A.~J.}\ \bibnamefont {Heinrich}},\ }\href {\doibase
  10.1103/PhysRevLett.114.076601} {\bibfield  {journal} {\bibinfo  {journal}
  {Phys. Rev. Lett.}\ }\textbf {\bibinfo {volume} {114}},\ \bibinfo {pages}
  {076601} (\bibinfo {year} {2015})}\BibitemShut {NoStop}%
\bibitem [{\citenamefont {Meierott}\ \emph {et~al.}(2015)\citenamefont
  {Meierott}, \citenamefont {N\'eel},\ and\ \citenamefont
  {Kr\"oger}}]{prb_91_201111}%
  \BibitemOpen
  \bibfield  {author} {\bibinfo {author} {\bibfnamefont {S.}~\bibnamefont
  {Meierott}}, \bibinfo {author} {\bibfnamefont {N.}~\bibnamefont {N\'eel}}, \
  and\ \bibinfo {author} {\bibfnamefont {J.}~\bibnamefont {Kr\"oger}},\ }\href
  {\doibase 10.1103/PhysRevB.91.201111} {\bibfield  {journal} {\bibinfo
  {journal} {Phys. Rev. B}\ }\textbf {\bibinfo {volume} {91}},\ \bibinfo
  {pages} {201111(R)} (\bibinfo {year} {2015})}\BibitemShut {NoStop}%
\bibitem [{\citenamefont {Wahl}\ \emph {et~al.}(2005)\citenamefont {Wahl},
  \citenamefont {Diekh\"oner}, \citenamefont {Wittich}, \citenamefont {Vitali},
  \citenamefont {Schneider},\ and\ \citenamefont {Kern}}]{prl_95_166601}%
  \BibitemOpen
  \bibfield  {author} {\bibinfo {author} {\bibfnamefont {P.}~\bibnamefont
  {Wahl}}, \bibinfo {author} {\bibfnamefont {L.}~\bibnamefont {Diekh\"oner}},
  \bibinfo {author} {\bibfnamefont {G.}~\bibnamefont {Wittich}}, \bibinfo
  {author} {\bibfnamefont {L.}~\bibnamefont {Vitali}}, \bibinfo {author}
  {\bibfnamefont {M.~A.}\ \bibnamefont {Schneider}}, \ and\ \bibinfo {author}
  {\bibfnamefont {K.}~\bibnamefont {Kern}},\ }\href {\doibase
  10.1103/PhysRevLett.95.166601} {\bibfield  {journal} {\bibinfo  {journal}
  {Phys. Rev. Lett.}\ }\textbf {\bibinfo {volume} {95}},\ \bibinfo {pages}
  {166601} (\bibinfo {year} {2005})}\BibitemShut {NoStop}%
\bibitem [{\citenamefont {Zhao}\ \emph {et~al.}(2005)\citenamefont {Zhao},
  \citenamefont {Li}, \citenamefont {Chen}, \citenamefont {Xiang},
  \citenamefont {Wang}, \citenamefont {Pan}, \citenamefont {Wang},
  \citenamefont {Xiao}, \citenamefont {Yang}, \citenamefont {Hou},\ and\
  \citenamefont {Zhu}}]{science_309_1542}%
  \BibitemOpen
  \bibfield  {author} {\bibinfo {author} {\bibfnamefont {A.}~\bibnamefont
  {Zhao}}, \bibinfo {author} {\bibfnamefont {Q.}~\bibnamefont {Li}}, \bibinfo
  {author} {\bibfnamefont {L.}~\bibnamefont {Chen}}, \bibinfo {author}
  {\bibfnamefont {H.}~\bibnamefont {Xiang}}, \bibinfo {author} {\bibfnamefont
  {W.}~\bibnamefont {Wang}}, \bibinfo {author} {\bibfnamefont {S.}~\bibnamefont
  {Pan}}, \bibinfo {author} {\bibfnamefont {B.}~\bibnamefont {Wang}}, \bibinfo
  {author} {\bibfnamefont {X.}~\bibnamefont {Xiao}}, \bibinfo {author}
  {\bibfnamefont {J.}~\bibnamefont {Yang}}, \bibinfo {author} {\bibfnamefont
  {J.~G.}\ \bibnamefont {Hou}}, \ and\ \bibinfo {author} {\bibfnamefont
  {Q.}~\bibnamefont {Zhu}},\ }\href {\doibase 10.1126/science.1113449}
  {\bibfield  {journal} {\bibinfo  {journal} {Science}\ }\textbf {\bibinfo
  {volume} {309}},\ \bibinfo {pages} {1542} (\bibinfo {year}
  {2005})}\BibitemShut {NoStop}%
\bibitem [{\citenamefont {Iancu}\ \emph
  {et~al.}(2006{\natexlab{a}})\citenamefont {Iancu}, \citenamefont
  {Deshpande},\ and\ \citenamefont {Hla}}]{prl_97_266603}%
  \BibitemOpen
  \bibfield  {author} {\bibinfo {author} {\bibfnamefont {V.}~\bibnamefont
  {Iancu}}, \bibinfo {author} {\bibfnamefont {A.}~\bibnamefont {Deshpande}}, \
  and\ \bibinfo {author} {\bibfnamefont {S.-W.}\ \bibnamefont {Hla}},\ }\href
  {\doibase 10.1103/PhysRevLett.97.266603} {\bibfield  {journal} {\bibinfo
  {journal} {Phys. Rev. Lett.}\ }\textbf {\bibinfo {volume} {97}},\ \bibinfo
  {pages} {266603} (\bibinfo {year} {2006}{\natexlab{a}})}\BibitemShut
  {NoStop}%
\bibitem [{\citenamefont {Iancu}\ \emph
  {et~al.}(2006{\natexlab{b}})\citenamefont {Iancu}, \citenamefont
  {Deshpande},\ and\ \citenamefont {Hla}}]{nl_6_820}%
  \BibitemOpen
  \bibfield  {author} {\bibinfo {author} {\bibfnamefont {V.}~\bibnamefont
  {Iancu}}, \bibinfo {author} {\bibfnamefont {A.}~\bibnamefont {Deshpande}}, \
  and\ \bibinfo {author} {\bibfnamefont {S.-W.}\ \bibnamefont {Hla}},\ }\href
  {\doibase 10.1021/nl0601886} {\bibfield  {journal} {\bibinfo  {journal} {Nano
  Letters}\ }\textbf {\bibinfo {volume} {6}},\ \bibinfo {pages} {820} (\bibinfo
  {year} {2006}{\natexlab{b}})}\BibitemShut {NoStop}%
\bibitem [{\citenamefont {Gao}\ \emph {et~al.}(2007)\citenamefont {Gao},
  \citenamefont {Ji}, \citenamefont {Hu}, \citenamefont {Cheng}, \citenamefont
  {Deng}, \citenamefont {Liu}, \citenamefont {Jiang}, \citenamefont {Lin},
  \citenamefont {Guo}, \citenamefont {Du}, \citenamefont {Hofer}, \citenamefont
  {Xie},\ and\ \citenamefont {Gao}}]{prl_99_106402}%
  \BibitemOpen
  \bibfield  {author} {\bibinfo {author} {\bibfnamefont {L.}~\bibnamefont
  {Gao}}, \bibinfo {author} {\bibfnamefont {W.}~\bibnamefont {Ji}}, \bibinfo
  {author} {\bibfnamefont {Y.~B.}\ \bibnamefont {Hu}}, \bibinfo {author}
  {\bibfnamefont {Z.~H.}\ \bibnamefont {Cheng}}, \bibinfo {author}
  {\bibfnamefont {Z.~T.}\ \bibnamefont {Deng}}, \bibinfo {author}
  {\bibfnamefont {Q.}~\bibnamefont {Liu}}, \bibinfo {author} {\bibfnamefont
  {N.}~\bibnamefont {Jiang}}, \bibinfo {author} {\bibfnamefont
  {X.}~\bibnamefont {Lin}}, \bibinfo {author} {\bibfnamefont {W.}~\bibnamefont
  {Guo}}, \bibinfo {author} {\bibfnamefont {S.~X.}\ \bibnamefont {Du}},
  \bibinfo {author} {\bibfnamefont {W.~A.}\ \bibnamefont {Hofer}}, \bibinfo
  {author} {\bibfnamefont {X.~C.}\ \bibnamefont {Xie}}, \ and\ \bibinfo
  {author} {\bibfnamefont {H.-J.}\ \bibnamefont {Gao}},\ }\href {\doibase
  10.1103/PhysRevLett.99.106402} {\bibfield  {journal} {\bibinfo  {journal}
  {Phys. Rev. Lett.}\ }\textbf {\bibinfo {volume} {99}},\ \bibinfo {pages}
  {106402} (\bibinfo {year} {2007})}\BibitemShut {NoStop}%
\bibitem [{\citenamefont {Fern\'andez-Torrente}\ \emph
  {et~al.}(2008)\citenamefont {Fern\'andez-Torrente}, \citenamefont {Franke},\
  and\ \citenamefont {Pascual}}]{prl_101_217203}%
  \BibitemOpen
  \bibfield  {author} {\bibinfo {author} {\bibfnamefont {I.}~\bibnamefont
  {Fern\'andez-Torrente}}, \bibinfo {author} {\bibfnamefont {K.~J.}\
  \bibnamefont {Franke}}, \ and\ \bibinfo {author} {\bibfnamefont {J.~I.}\
  \bibnamefont {Pascual}},\ }\href {\doibase 10.1103/PhysRevLett.101.217203}
  {\bibfield  {journal} {\bibinfo  {journal} {Phys. Rev. Lett.}\ }\textbf
  {\bibinfo {volume} {101}},\ \bibinfo {pages} {217203} (\bibinfo {year}
  {2008})}\BibitemShut {NoStop}%
\bibitem [{\citenamefont {Perera}\ \emph {et~al.}(2010)\citenamefont {Perera},
  \citenamefont {Kulik}, \citenamefont {Iancu}, \citenamefont {Dias~da Silva},
  \citenamefont {Ulloa}, \citenamefont {Marzari},\ and\ \citenamefont
  {Hla}}]{prl_105_106601}%
  \BibitemOpen
  \bibfield  {author} {\bibinfo {author} {\bibfnamefont {U.~G.~E.}\
  \bibnamefont {Perera}}, \bibinfo {author} {\bibfnamefont {H.~J.}\
  \bibnamefont {Kulik}}, \bibinfo {author} {\bibfnamefont {V.}~\bibnamefont
  {Iancu}}, \bibinfo {author} {\bibfnamefont {L.~G. G.~V.}\ \bibnamefont
  {Dias~da Silva}}, \bibinfo {author} {\bibfnamefont {S.~E.}\ \bibnamefont
  {Ulloa}}, \bibinfo {author} {\bibfnamefont {N.}~\bibnamefont {Marzari}}, \
  and\ \bibinfo {author} {\bibfnamefont {S.-W.}\ \bibnamefont {Hla}},\ }\href
  {\doibase 10.1103/PhysRevLett.105.106601} {\bibfield  {journal} {\bibinfo
  {journal} {Phys. Rev. Lett.}\ }\textbf {\bibinfo {volume} {105}},\ \bibinfo
  {pages} {106601} (\bibinfo {year} {2010})}\BibitemShut {NoStop}%
\bibitem [{\citenamefont {Choi}\ \emph {et~al.}(2010)\citenamefont {Choi},
  \citenamefont {Bedwani}, \citenamefont {Rochefort}, \citenamefont {Chen},
  \citenamefont {Epstein},\ and\ \citenamefont {Gupta}}]{nl_10_4175}%
  \BibitemOpen
  \bibfield  {author} {\bibinfo {author} {\bibfnamefont {T.}~\bibnamefont
  {Choi}}, \bibinfo {author} {\bibfnamefont {S.}~\bibnamefont {Bedwani}},
  \bibinfo {author} {\bibfnamefont {A.}~\bibnamefont {Rochefort}}, \bibinfo
  {author} {\bibfnamefont {C.-Y.}\ \bibnamefont {Chen}}, \bibinfo {author}
  {\bibfnamefont {A.~J.}\ \bibnamefont {Epstein}}, \ and\ \bibinfo {author}
  {\bibfnamefont {J.~A.}\ \bibnamefont {Gupta}},\ }\href {\doibase
  10.1021/nl1024563} {\bibfield  {journal} {\bibinfo  {journal} {Nano Letters}\
  }\textbf {\bibinfo {volume} {10}},\ \bibinfo {pages} {4175} (\bibinfo {year}
  {2010})}\BibitemShut {NoStop}%
\bibitem [{\citenamefont {Komeda}\ \emph {et~al.}(2011)\citenamefont {Komeda},
  \citenamefont {Isshiki}, \citenamefont {Liu}, \citenamefont {Zhang},
  \citenamefont {Lorente}, \citenamefont {Katoh}, \citenamefont {Breedlove},\
  and\ \citenamefont {Yamashita}}]{natcommun_2_217}%
  \BibitemOpen
  \bibfield  {author} {\bibinfo {author} {\bibfnamefont {T.}~\bibnamefont
  {Komeda}}, \bibinfo {author} {\bibfnamefont {H.}~\bibnamefont {Isshiki}},
  \bibinfo {author} {\bibfnamefont {J.}~\bibnamefont {Liu}}, \bibinfo {author}
  {\bibfnamefont {Y.-F.}\ \bibnamefont {Zhang}}, \bibinfo {author}
  {\bibfnamefont {N.}~\bibnamefont {Lorente}}, \bibinfo {author} {\bibfnamefont
  {K.}~\bibnamefont {Katoh}}, \bibinfo {author} {\bibfnamefont {B.~K.}\
  \bibnamefont {Breedlove}}, \ and\ \bibinfo {author} {\bibfnamefont
  {M.}~\bibnamefont {Yamashita}},\ }\href {\doibase 10.1038/ncomms1210}
  {\bibfield  {journal} {\bibinfo  {journal} {Nature Communications}\ }\textbf
  {\bibinfo {volume} {2}},\ \bibinfo {pages} {217} (\bibinfo {year}
  {2011})}\BibitemShut {NoStop}%
\bibitem [{\citenamefont {Mugarza}\ \emph {et~al.}(2011)\citenamefont
  {Mugarza}, \citenamefont {Krull}, \citenamefont {Robles}, \citenamefont
  {Stepanow}, \citenamefont {Ceballos},\ and\ \citenamefont
  {Gambardella}}]{natcommun_2_490}%
  \BibitemOpen
  \bibfield  {author} {\bibinfo {author} {\bibfnamefont {A.}~\bibnamefont
  {Mugarza}}, \bibinfo {author} {\bibfnamefont {C.}~\bibnamefont {Krull}},
  \bibinfo {author} {\bibfnamefont {R.}~\bibnamefont {Robles}}, \bibinfo
  {author} {\bibfnamefont {S.}~\bibnamefont {Stepanow}}, \bibinfo {author}
  {\bibfnamefont {G.}~\bibnamefont {Ceballos}}, \ and\ \bibinfo {author}
  {\bibfnamefont {P.}~\bibnamefont {Gambardella}},\ }\href {\doibase
  10.1038/ncomms1497} {\bibfield  {journal} {\bibinfo  {journal} {Nature
  Communications}\ }\textbf {\bibinfo {volume} {2}},\ \bibinfo {pages} {490}
  (\bibinfo {year} {2011})}\BibitemShut {NoStop}%
\bibitem [{\citenamefont {Tsukahara}\ \emph {et~al.}(2011)\citenamefont
  {Tsukahara}, \citenamefont {Shiraki}, \citenamefont {Itou}, \citenamefont
  {Ohta}, \citenamefont {Takagi},\ and\ \citenamefont
  {Kawai}}]{prl_106_187201}%
  \BibitemOpen
  \bibfield  {author} {\bibinfo {author} {\bibfnamefont {N.}~\bibnamefont
  {Tsukahara}}, \bibinfo {author} {\bibfnamefont {S.}~\bibnamefont {Shiraki}},
  \bibinfo {author} {\bibfnamefont {S.}~\bibnamefont {Itou}}, \bibinfo {author}
  {\bibfnamefont {N.}~\bibnamefont {Ohta}}, \bibinfo {author} {\bibfnamefont
  {N.}~\bibnamefont {Takagi}}, \ and\ \bibinfo {author} {\bibfnamefont
  {M.}~\bibnamefont {Kawai}},\ }\href {\doibase 10.1103/PhysRevLett.106.187201}
  {\bibfield  {journal} {\bibinfo  {journal} {Phys. Rev. Lett.}\ }\textbf
  {\bibinfo {volume} {106}},\ \bibinfo {pages} {187201} (\bibinfo {year}
  {2011})}\BibitemShut {NoStop}%
\bibitem [{\citenamefont {Franke}\ \emph {et~al.}(2011)\citenamefont {Franke},
  \citenamefont {Schulze},\ and\ \citenamefont {Pascual}}]{science_332_940}%
  \BibitemOpen
  \bibfield  {author} {\bibinfo {author} {\bibfnamefont {K.~J.}\ \bibnamefont
  {Franke}}, \bibinfo {author} {\bibfnamefont {G.}~\bibnamefont {Schulze}}, \
  and\ \bibinfo {author} {\bibfnamefont {J.~I.}\ \bibnamefont {Pascual}},\
  }\href {\doibase 10.1126/science.1202204} {\bibfield  {journal} {\bibinfo
  {journal} {Science}\ }\textbf {\bibinfo {volume} {332}},\ \bibinfo {pages}
  {940} (\bibinfo {year} {2011})}\BibitemShut {NoStop}%
\bibitem [{\citenamefont {DiLullo}\ \emph {et~al.}(2012)\citenamefont
  {DiLullo}, \citenamefont {Chang}, \citenamefont {Baadji}, \citenamefont
  {Clark}, \citenamefont {Klöckner}, \citenamefont {Prosenc}, \citenamefont
  {Sanvito}, \citenamefont {Wiesendanger}, \citenamefont {Hoffmann},\ and\
  \citenamefont {Hla}}]{nl_12_3174}%
  \BibitemOpen
  \bibfield  {author} {\bibinfo {author} {\bibfnamefont {A.}~\bibnamefont
  {DiLullo}}, \bibinfo {author} {\bibfnamefont {S.-H.}\ \bibnamefont {Chang}},
  \bibinfo {author} {\bibfnamefont {N.}~\bibnamefont {Baadji}}, \bibinfo
  {author} {\bibfnamefont {K.}~\bibnamefont {Clark}}, \bibinfo {author}
  {\bibfnamefont {J.-P.}\ \bibnamefont {Klöckner}}, \bibinfo {author}
  {\bibfnamefont {M.-H.}\ \bibnamefont {Prosenc}}, \bibinfo {author}
  {\bibfnamefont {S.}~\bibnamefont {Sanvito}}, \bibinfo {author} {\bibfnamefont
  {R.}~\bibnamefont {Wiesendanger}}, \bibinfo {author} {\bibfnamefont
  {G.}~\bibnamefont {Hoffmann}}, \ and\ \bibinfo {author} {\bibfnamefont
  {S.-W.}\ \bibnamefont {Hla}},\ }\href {\doibase 10.1021/nl301149d} {\bibfield
   {journal} {\bibinfo  {journal} {Nano Letters}\ }\textbf {\bibinfo {volume}
  {12}},\ \bibinfo {pages} {3174} (\bibinfo {year} {2012})}\BibitemShut
  {NoStop}%
\bibitem [{\citenamefont {Robles}\ \emph {et~al.}(2012)\citenamefont {Robles},
  \citenamefont {Lorente}, \citenamefont {Isshiki}, \citenamefont {Liu},
  \citenamefont {Katoh}, \citenamefont {Breedlove}, \citenamefont {Yamashita},\
  and\ \citenamefont {Komeda}}]{nl_12_3609}%
  \BibitemOpen
  \bibfield  {author} {\bibinfo {author} {\bibfnamefont {R.}~\bibnamefont
  {Robles}}, \bibinfo {author} {\bibfnamefont {N.}~\bibnamefont {Lorente}},
  \bibinfo {author} {\bibfnamefont {H.}~\bibnamefont {Isshiki}}, \bibinfo
  {author} {\bibfnamefont {J.}~\bibnamefont {Liu}}, \bibinfo {author}
  {\bibfnamefont {K.}~\bibnamefont {Katoh}}, \bibinfo {author} {\bibfnamefont
  {B.~K.}\ \bibnamefont {Breedlove}}, \bibinfo {author} {\bibfnamefont
  {M.}~\bibnamefont {Yamashita}}, \ and\ \bibinfo {author} {\bibfnamefont
  {T.}~\bibnamefont {Komeda}},\ }\href {\doibase 10.1021/nl301301e} {\bibfield
  {journal} {\bibinfo  {journal} {Nano Letters}\ }\textbf {\bibinfo {volume}
  {12}},\ \bibinfo {pages} {3609} (\bibinfo {year} {2012})}\BibitemShut
  {NoStop}%
\bibitem [{\citenamefont {Gopakumar}\ \emph {et~al.}(2012)\citenamefont
  {Gopakumar}, \citenamefont {Matino}, \citenamefont {Naggert}, \citenamefont
  {Bannwarth}, \citenamefont {Tuczek},\ and\ \citenamefont
  {Berndt}}]{acie_51_6262}%
  \BibitemOpen
  \bibfield  {author} {\bibinfo {author} {\bibfnamefont {T.~G.}\ \bibnamefont
  {Gopakumar}}, \bibinfo {author} {\bibfnamefont {F.}~\bibnamefont {Matino}},
  \bibinfo {author} {\bibfnamefont {H.}~\bibnamefont {Naggert}}, \bibinfo
  {author} {\bibfnamefont {A.}~\bibnamefont {Bannwarth}}, \bibinfo {author}
  {\bibfnamefont {F.}~\bibnamefont {Tuczek}}, \ and\ \bibinfo {author}
  {\bibfnamefont {R.}~\bibnamefont {Berndt}},\ }\href {\doibase
  10.1002/anie.201201203} {\bibfield  {journal} {\bibinfo  {journal}
  {Angewandte Chemie International Edition}\ }\textbf {\bibinfo {volume}
  {51}},\ \bibinfo {pages} {6262} (\bibinfo {year} {2012})}\BibitemShut
  {NoStop}%
\bibitem [{\citenamefont {Miyamachi}\ \emph {et~al.}(2012)\citenamefont
  {Miyamachi}, \citenamefont {Gruber}, \citenamefont {Davesne}, \citenamefont
  {Bowen}, \citenamefont {Boukari}, \citenamefont {Joly}, \citenamefont
  {Scheurer}, \citenamefont {Rogez}, \citenamefont {Yamada}, \citenamefont
  {Ohresser}, \citenamefont {Beaurepaire},\ and\ \citenamefont
  {Wulfhekel}}]{natcommun_3_938}%
  \BibitemOpen
  \bibfield  {author} {\bibinfo {author} {\bibfnamefont {T.}~\bibnamefont
  {Miyamachi}}, \bibinfo {author} {\bibfnamefont {M.}~\bibnamefont {Gruber}},
  \bibinfo {author} {\bibfnamefont {V.}~\bibnamefont {Davesne}}, \bibinfo
  {author} {\bibfnamefont {M.}~\bibnamefont {Bowen}}, \bibinfo {author}
  {\bibfnamefont {S.}~\bibnamefont {Boukari}}, \bibinfo {author} {\bibfnamefont
  {L.}~\bibnamefont {Joly}}, \bibinfo {author} {\bibfnamefont {F.}~\bibnamefont
  {Scheurer}}, \bibinfo {author} {\bibfnamefont {G.}~\bibnamefont {Rogez}},
  \bibinfo {author} {\bibfnamefont {T.~K.}\ \bibnamefont {Yamada}}, \bibinfo
  {author} {\bibfnamefont {P.}~\bibnamefont {Ohresser}}, \bibinfo {author}
  {\bibfnamefont {E.}~\bibnamefont {Beaurepaire}}, \ and\ \bibinfo {author}
  {\bibfnamefont {W.}~\bibnamefont {Wulfhekel}},\ }\href {\doibase
  10.1038/ncomms1940} {\bibfield  {journal} {\bibinfo  {journal} {Nature
  Communications}\ }\textbf {\bibinfo {volume} {3}},\ \bibinfo {pages} {938}
  (\bibinfo {year} {2012})}\BibitemShut {NoStop}%
\bibitem [{\citenamefont {Minamitani}\ \emph {et~al.}(2012)\citenamefont
  {Minamitani}, \citenamefont {Tsukahara}, \citenamefont {Matsunaka},
  \citenamefont {Kim}, \citenamefont {Takagi},\ and\ \citenamefont
  {Kawai}}]{prl_109_086602}%
  \BibitemOpen
  \bibfield  {author} {\bibinfo {author} {\bibfnamefont {E.}~\bibnamefont
  {Minamitani}}, \bibinfo {author} {\bibfnamefont {N.}~\bibnamefont
  {Tsukahara}}, \bibinfo {author} {\bibfnamefont {D.}~\bibnamefont
  {Matsunaka}}, \bibinfo {author} {\bibfnamefont {Y.}~\bibnamefont {Kim}},
  \bibinfo {author} {\bibfnamefont {N.}~\bibnamefont {Takagi}}, \ and\ \bibinfo
  {author} {\bibfnamefont {M.}~\bibnamefont {Kawai}},\ }\href {\doibase
  10.1103/PhysRevLett.109.086602} {\bibfield  {journal} {\bibinfo  {journal}
  {Phys. Rev. Lett.}\ }\textbf {\bibinfo {volume} {109}},\ \bibinfo {pages}
  {086602} (\bibinfo {year} {2012})}\BibitemShut {NoStop}%
\bibitem [{\citenamefont {Kim}\ \emph {et~al.}(2013)\citenamefont {Kim},
  \citenamefont {Chang}, \citenamefont {Lee}, \citenamefont {Kim},\ and\
  \citenamefont {Kahng}}]{acsnano_7_9312}%
  \BibitemOpen
  \bibfield  {author} {\bibinfo {author} {\bibfnamefont {H.}~\bibnamefont
  {Kim}}, \bibinfo {author} {\bibfnamefont {Y.~H.}\ \bibnamefont {Chang}},
  \bibinfo {author} {\bibfnamefont {S.-H.}\ \bibnamefont {Lee}}, \bibinfo
  {author} {\bibfnamefont {Y.-H.}\ \bibnamefont {Kim}}, \ and\ \bibinfo
  {author} {\bibfnamefont {S.-J.}\ \bibnamefont {Kahng}},\ }\href {\doibase
  10.1021/nn4039595} {\bibfield  {journal} {\bibinfo  {journal} {ACS Nano}\
  }\textbf {\bibinfo {volume} {7}},\ \bibinfo {pages} {9312} (\bibinfo {year}
  {2013})}\BibitemShut {NoStop}%
\bibitem [{\citenamefont {Heinrich}\ \emph {et~al.}(2013)\citenamefont
  {Heinrich}, \citenamefont {Ahmadi}, \citenamefont {Müller}, \citenamefont
  {Braun}, \citenamefont {Pascual},\ and\ \citenamefont {Franke}}]{nl_13_4840}%
  \BibitemOpen
  \bibfield  {author} {\bibinfo {author} {\bibfnamefont {B.~W.}\ \bibnamefont
  {Heinrich}}, \bibinfo {author} {\bibfnamefont {G.}~\bibnamefont {Ahmadi}},
  \bibinfo {author} {\bibfnamefont {V.~L.}\ \bibnamefont {Müller}}, \bibinfo
  {author} {\bibfnamefont {L.}~\bibnamefont {Braun}}, \bibinfo {author}
  {\bibfnamefont {J.~I.}\ \bibnamefont {Pascual}}, \ and\ \bibinfo {author}
  {\bibfnamefont {K.~J.}\ \bibnamefont {Franke}},\ }\href {\doibase
  10.1021/nl402575c} {\bibfield  {journal} {\bibinfo  {journal} {Nano Letters}\
  }\textbf {\bibinfo {volume} {13}},\ \bibinfo {pages} {4840} (\bibinfo {year}
  {2013})}\BibitemShut {NoStop}%
\bibitem [{\citenamefont {Ternes}\ \emph {et~al.}(2008)\citenamefont {Ternes},
  \citenamefont {Heinrich},\ and\ \citenamefont {Schneider}}]{jpcm_21_053001}%
  \BibitemOpen
  \bibfield  {author} {\bibinfo {author} {\bibfnamefont {M.}~\bibnamefont
  {Ternes}}, \bibinfo {author} {\bibfnamefont {A.~J.}\ \bibnamefont
  {Heinrich}}, \ and\ \bibinfo {author} {\bibfnamefont {W.-D.}\ \bibnamefont
  {Schneider}},\ }\href {\doibase 10.1088/0953-8984/21/5/053001} {\bibfield
  {journal} {\bibinfo  {journal} {Journal of Physics: Condensed Matter}\
  }\textbf {\bibinfo {volume} {21}},\ \bibinfo {pages} {053001} (\bibinfo
  {year} {2008})}\BibitemShut {NoStop}%
\bibitem [{\citenamefont {Komeda}(2014)}]{ss_630_343}%
  \BibitemOpen
  \bibfield  {author} {\bibinfo {author} {\bibfnamefont {T.}~\bibnamefont
  {Komeda}},\ }\href {\doibase https://doi.org/10.1016/j.susc.2014.07.012}
  {\bibfield  {journal} {\bibinfo  {journal} {Surface Science}\ }\textbf
  {\bibinfo {volume} {630}},\ \bibinfo {pages} {343 } (\bibinfo {year}
  {2014})}\BibitemShut {NoStop}%
\bibitem [{\citenamefont {Ternes}(2017)}]{pss_92_83}%
  \BibitemOpen
  \bibfield  {author} {\bibinfo {author} {\bibfnamefont {M.}~\bibnamefont
  {Ternes}},\ }\href {\doibase https://doi.org/10.1016/j.progsurf.2017.01.001}
  {\bibfield  {journal} {\bibinfo  {journal} {Progress in Surface Science}\
  }\textbf {\bibinfo {volume} {92}},\ \bibinfo {pages} {83 } (\bibinfo {year}
  {2017})}\BibitemShut {NoStop}%
\bibitem [{\citenamefont {Chen}\ \emph {et~al.}(1999)\citenamefont {Chen},
  \citenamefont {Jamneala}, \citenamefont {Madhavan},\ and\ \citenamefont
  {Crommie}}]{prb_60_r8529}%
  \BibitemOpen
  \bibfield  {author} {\bibinfo {author} {\bibfnamefont {W.}~\bibnamefont
  {Chen}}, \bibinfo {author} {\bibfnamefont {T.}~\bibnamefont {Jamneala}},
  \bibinfo {author} {\bibfnamefont {V.}~\bibnamefont {Madhavan}}, \ and\
  \bibinfo {author} {\bibfnamefont {M.~F.}\ \bibnamefont {Crommie}},\ }\href
  {\doibase 10.1103/PhysRevB.60.R8529} {\bibfield  {journal} {\bibinfo
  {journal} {Phys. Rev. B}\ }\textbf {\bibinfo {volume} {60}},\ \bibinfo
  {pages} {R8529} (\bibinfo {year} {1999})}\BibitemShut {NoStop}%
\bibitem [{\citenamefont {Madhavan}\ \emph {et~al.}(2002)\citenamefont
  {Madhavan}, \citenamefont {Jamneala}, \citenamefont {Nagaoka}, \citenamefont
  {Chen}, \citenamefont {Li}, \citenamefont {Louie},\ and\ \citenamefont
  {Crommie}}]{prb_66_212411}%
  \BibitemOpen
  \bibfield  {author} {\bibinfo {author} {\bibfnamefont {V.}~\bibnamefont
  {Madhavan}}, \bibinfo {author} {\bibfnamefont {T.}~\bibnamefont {Jamneala}},
  \bibinfo {author} {\bibfnamefont {K.}~\bibnamefont {Nagaoka}}, \bibinfo
  {author} {\bibfnamefont {W.}~\bibnamefont {Chen}}, \bibinfo {author}
  {\bibfnamefont {J.-L.}\ \bibnamefont {Li}}, \bibinfo {author} {\bibfnamefont
  {S.~G.}\ \bibnamefont {Louie}}, \ and\ \bibinfo {author} {\bibfnamefont
  {M.~F.}\ \bibnamefont {Crommie}},\ }\href {\doibase
  10.1103/PhysRevB.66.212411} {\bibfield  {journal} {\bibinfo  {journal} {Phys.
  Rev. B}\ }\textbf {\bibinfo {volume} {66}},\ \bibinfo {pages} {212411}
  (\bibinfo {year} {2002})}\BibitemShut {NoStop}%
\bibitem [{\citenamefont {Wahl}\ \emph {et~al.}(2007)\citenamefont {Wahl},
  \citenamefont {Simon}, \citenamefont {Diekh\"oner}, \citenamefont
  {Stepanyuk}, \citenamefont {Bruno}, \citenamefont {Schneider},\ and\
  \citenamefont {Kern}}]{prl_98_056601}%
  \BibitemOpen
  \bibfield  {author} {\bibinfo {author} {\bibfnamefont {P.}~\bibnamefont
  {Wahl}}, \bibinfo {author} {\bibfnamefont {P.}~\bibnamefont {Simon}},
  \bibinfo {author} {\bibfnamefont {L.}~\bibnamefont {Diekh\"oner}}, \bibinfo
  {author} {\bibfnamefont {V.~S.}\ \bibnamefont {Stepanyuk}}, \bibinfo {author}
  {\bibfnamefont {P.}~\bibnamefont {Bruno}}, \bibinfo {author} {\bibfnamefont
  {M.~A.}\ \bibnamefont {Schneider}}, \ and\ \bibinfo {author} {\bibfnamefont
  {K.}~\bibnamefont {Kern}},\ }\href {\doibase 10.1103/PhysRevLett.98.056601}
  {\bibfield  {journal} {\bibinfo  {journal} {Phys. Rev. Lett.}\ }\textbf
  {\bibinfo {volume} {98}},\ \bibinfo {pages} {056601} (\bibinfo {year}
  {2007})}\BibitemShut {NoStop}%
\bibitem [{\citenamefont {Otte}\ \emph {et~al.}(2008)\citenamefont {Otte},
  \citenamefont {Ternes}, \citenamefont {von Bergmann}, \citenamefont {Loth},
  \citenamefont {Brune}, \citenamefont {Lutz}, \citenamefont {Hirjibehedin},\
  and\ \citenamefont {Heinrich}}]{natphys_4_847}%
  \BibitemOpen
  \bibfield  {author} {\bibinfo {author} {\bibfnamefont {A.~F.}\ \bibnamefont
  {Otte}}, \bibinfo {author} {\bibfnamefont {M.}~\bibnamefont {Ternes}},
  \bibinfo {author} {\bibfnamefont {K.}~\bibnamefont {von Bergmann}}, \bibinfo
  {author} {\bibfnamefont {S.}~\bibnamefont {Loth}}, \bibinfo {author}
  {\bibfnamefont {H.}~\bibnamefont {Brune}}, \bibinfo {author} {\bibfnamefont
  {C.~P.}\ \bibnamefont {Lutz}}, \bibinfo {author} {\bibfnamefont {C.~F.}\
  \bibnamefont {Hirjibehedin}}, \ and\ \bibinfo {author} {\bibfnamefont
  {A.~J.}\ \bibnamefont {Heinrich}},\ }\href {\doibase 10.1038/nphys1072}
  {\bibfield  {journal} {\bibinfo  {journal} {Nature Physics}\ }\textbf
  {\bibinfo {volume} {4}},\ \bibinfo {pages} {847 } (\bibinfo {year}
  {2008})}\BibitemShut {NoStop}%
\bibitem [{\citenamefont {Otte}\ \emph {et~al.}(2009)\citenamefont {Otte},
  \citenamefont {Ternes}, \citenamefont {Loth}, \citenamefont {Lutz},
  \citenamefont {Hirjibehedin},\ and\ \citenamefont
  {Heinrich}}]{prl_103_107203}%
  \BibitemOpen
  \bibfield  {author} {\bibinfo {author} {\bibfnamefont {A.~F.}\ \bibnamefont
  {Otte}}, \bibinfo {author} {\bibfnamefont {M.}~\bibnamefont {Ternes}},
  \bibinfo {author} {\bibfnamefont {S.}~\bibnamefont {Loth}}, \bibinfo {author}
  {\bibfnamefont {C.~P.}\ \bibnamefont {Lutz}}, \bibinfo {author}
  {\bibfnamefont {C.~F.}\ \bibnamefont {Hirjibehedin}}, \ and\ \bibinfo
  {author} {\bibfnamefont {A.~J.}\ \bibnamefont {Heinrich}},\ }\href {\doibase
  10.1103/PhysRevLett.103.107203} {\bibfield  {journal} {\bibinfo  {journal}
  {Phys. Rev. Lett.}\ }\textbf {\bibinfo {volume} {103}},\ \bibinfo {pages}
  {107203} (\bibinfo {year} {2009})}\BibitemShut {NoStop}%
\bibitem [{\citenamefont {N\'eel}\ \emph {et~al.}(2011)\citenamefont {N\'eel},
  \citenamefont {Berndt}, \citenamefont {Kr\"oger}, \citenamefont {Wehling},
  \citenamefont {Lichtenstein},\ and\ \citenamefont
  {Katsnelson}}]{prl_107_106804}%
  \BibitemOpen
  \bibfield  {author} {\bibinfo {author} {\bibfnamefont {N.}~\bibnamefont
  {N\'eel}}, \bibinfo {author} {\bibfnamefont {R.}~\bibnamefont {Berndt}},
  \bibinfo {author} {\bibfnamefont {J.}~\bibnamefont {Kr\"oger}}, \bibinfo
  {author} {\bibfnamefont {T.~O.}\ \bibnamefont {Wehling}}, \bibinfo {author}
  {\bibfnamefont {A.~I.}\ \bibnamefont {Lichtenstein}}, \ and\ \bibinfo
  {author} {\bibfnamefont {M.~I.}\ \bibnamefont {Katsnelson}},\ }\href
  {\doibase 10.1103/PhysRevLett.107.106804} {\bibfield  {journal} {\bibinfo
  {journal} {Phys. Rev. Lett.}\ }\textbf {\bibinfo {volume} {107}},\ \bibinfo
  {pages} {106804} (\bibinfo {year} {2011})}\BibitemShut {NoStop}%
\bibitem [{\citenamefont {Bork}\ \emph {et~al.}(2011)\citenamefont {Bork},
  \citenamefont {Zhang}, \citenamefont {Diekhöner}, \citenamefont {Borda},
  \citenamefont {Simon}, \citenamefont {Kroha}, \citenamefont {Wahl},\ and\
  \citenamefont {Kern}}]{natphys_7_901}%
  \BibitemOpen
  \bibfield  {author} {\bibinfo {author} {\bibfnamefont {J.}~\bibnamefont
  {Bork}}, \bibinfo {author} {\bibfnamefont {Y.-h.}\ \bibnamefont {Zhang}},
  \bibinfo {author} {\bibfnamefont {L.}~\bibnamefont {Diekhöner}}, \bibinfo
  {author} {\bibfnamefont {L.}~\bibnamefont {Borda}}, \bibinfo {author}
  {\bibfnamefont {P.}~\bibnamefont {Simon}}, \bibinfo {author} {\bibfnamefont
  {J.}~\bibnamefont {Kroha}}, \bibinfo {author} {\bibfnamefont
  {P.}~\bibnamefont {Wahl}}, \ and\ \bibinfo {author} {\bibfnamefont
  {K.}~\bibnamefont {Kern}},\ }\href {\doibase 10.1038/nphys2076} {\bibfield
  {journal} {\bibinfo  {journal} {Nature Physics}\ }\textbf {\bibinfo {volume}
  {7}},\ \bibinfo {pages} {901 } (\bibinfo {year} {2011})}\BibitemShut
  {NoStop}%
\bibitem [{\citenamefont {Prüser}\ \emph {et~al.}(2014)\citenamefont
  {Prüser}, \citenamefont {Dargel}, \citenamefont {Bouhassoune}, \citenamefont
  {Ulbrich}, \citenamefont {Pruschke}, \citenamefont {Lounis},\ and\
  \citenamefont {Wenderoth}}]{natcommun_5_5417}%
  \BibitemOpen
  \bibfield  {author} {\bibinfo {author} {\bibfnamefont {H.}~\bibnamefont
  {Prüser}}, \bibinfo {author} {\bibfnamefont {P.~E.}\ \bibnamefont {Dargel}},
  \bibinfo {author} {\bibfnamefont {M.}~\bibnamefont {Bouhassoune}}, \bibinfo
  {author} {\bibfnamefont {R.~G.}\ \bibnamefont {Ulbrich}}, \bibinfo {author}
  {\bibfnamefont {T.}~\bibnamefont {Pruschke}}, \bibinfo {author}
  {\bibfnamefont {S.}~\bibnamefont {Lounis}}, \ and\ \bibinfo {author}
  {\bibfnamefont {M.}~\bibnamefont {Wenderoth}},\ }\href {\doibase
  10.1038/ncomms6417} {\bibfield  {journal} {\bibinfo  {journal} {Nature
  Communications}\ }\textbf {\bibinfo {volume} {5}},\ \bibinfo {pages} {5417}
  (\bibinfo {year} {2014})}\BibitemShut {NoStop}%
\bibitem [{\citenamefont {Spinelli}\ \emph {et~al.}(2015)\citenamefont
  {Spinelli}, \citenamefont {Gerrits}, \citenamefont {Toskovic}, \citenamefont
  {Bryant}, \citenamefont {Ternes},\ and\ \citenamefont
  {Otte}}]{natcommun_6_10046}%
  \BibitemOpen
  \bibfield  {author} {\bibinfo {author} {\bibfnamefont {A.}~\bibnamefont
  {Spinelli}}, \bibinfo {author} {\bibfnamefont {M.}~\bibnamefont {Gerrits}},
  \bibinfo {author} {\bibfnamefont {R.}~\bibnamefont {Toskovic}}, \bibinfo
  {author} {\bibfnamefont {B.}~\bibnamefont {Bryant}}, \bibinfo {author}
  {\bibfnamefont {M.}~\bibnamefont {Ternes}}, \ and\ \bibinfo {author}
  {\bibfnamefont {A.~F.}\ \bibnamefont {Otte}},\ }\href {\doibase
  10.1038/ncomms10046} {\bibfield  {journal} {\bibinfo  {journal} {Nature
  Communications}\ }\textbf {\bibinfo {volume} {6}},\ \bibinfo {pages} {10046}
  (\bibinfo {year} {2015})}\BibitemShut {NoStop}%
\bibitem [{\citenamefont {Choi}\ \emph {et~al.}(2016)\citenamefont {Choi},
  \citenamefont {Guissart}, \citenamefont {Ormaza}, \citenamefont {Bachellier},
  \citenamefont {Bengone}, \citenamefont {Simon},\ and\ \citenamefont
  {Limot}}]{nl_16_6298}%
  \BibitemOpen
  \bibfield  {author} {\bibinfo {author} {\bibfnamefont {D.-J.}\ \bibnamefont
  {Choi}}, \bibinfo {author} {\bibfnamefont {S.}~\bibnamefont {Guissart}},
  \bibinfo {author} {\bibfnamefont {M.}~\bibnamefont {Ormaza}}, \bibinfo
  {author} {\bibfnamefont {N.}~\bibnamefont {Bachellier}}, \bibinfo {author}
  {\bibfnamefont {O.}~\bibnamefont {Bengone}}, \bibinfo {author} {\bibfnamefont
  {P.}~\bibnamefont {Simon}}, \ and\ \bibinfo {author} {\bibfnamefont
  {L.}~\bibnamefont {Limot}},\ }\href {\doibase 10.1021/acs.nanolett.6b02617}
  {\bibfield  {journal} {\bibinfo  {journal} {Nano Letters}\ }\textbf {\bibinfo
  {volume} {16}},\ \bibinfo {pages} {6298} (\bibinfo {year}
  {2016})}\BibitemShut {NoStop}%
\bibitem [{\citenamefont {N\'eel}\ \emph {et~al.}(2007)\citenamefont {N\'eel},
  \citenamefont {Kr\"oger}, \citenamefont {Limot}, \citenamefont {Palotas},
  \citenamefont {Hofer},\ and\ \citenamefont {Berndt}}]{prl_98_016801}%
  \BibitemOpen
  \bibfield  {author} {\bibinfo {author} {\bibfnamefont {N.}~\bibnamefont
  {N\'eel}}, \bibinfo {author} {\bibfnamefont {J.}~\bibnamefont {Kr\"oger}},
  \bibinfo {author} {\bibfnamefont {L.}~\bibnamefont {Limot}}, \bibinfo
  {author} {\bibfnamefont {K.}~\bibnamefont {Palotas}}, \bibinfo {author}
  {\bibfnamefont {W.~A.}\ \bibnamefont {Hofer}}, \ and\ \bibinfo {author}
  {\bibfnamefont {R.}~\bibnamefont {Berndt}},\ }\href {\doibase
  10.1103/PhysRevLett.98.016801} {\bibfield  {journal} {\bibinfo  {journal}
  {Phys. Rev. Lett.}\ }\textbf {\bibinfo {volume} {98}},\ \bibinfo {pages}
  {016801} (\bibinfo {year} {2007})}\BibitemShut {NoStop}%
\bibitem [{\citenamefont {N\'eel}\ \emph {et~al.}(2008)\citenamefont {N\'eel},
  \citenamefont {Kr\"oger}, \citenamefont {Berndt}, \citenamefont {Wehling},
  \citenamefont {Lichtenstein},\ and\ \citenamefont
  {Katsnelson}}]{prl_101_266803}%
  \BibitemOpen
  \bibfield  {author} {\bibinfo {author} {\bibfnamefont {N.}~\bibnamefont
  {N\'eel}}, \bibinfo {author} {\bibfnamefont {J.}~\bibnamefont {Kr\"oger}},
  \bibinfo {author} {\bibfnamefont {R.}~\bibnamefont {Berndt}}, \bibinfo
  {author} {\bibfnamefont {T.~O.}\ \bibnamefont {Wehling}}, \bibinfo {author}
  {\bibfnamefont {A.~I.}\ \bibnamefont {Lichtenstein}}, \ and\ \bibinfo
  {author} {\bibfnamefont {M.~I.}\ \bibnamefont {Katsnelson}},\ }\href
  {\doibase 10.1103/PhysRevLett.101.266803} {\bibfield  {journal} {\bibinfo
  {journal} {Phys. Rev. Lett.}\ }\textbf {\bibinfo {volume} {101}},\ \bibinfo
  {pages} {266803} (\bibinfo {year} {2008})}\BibitemShut {NoStop}%
\bibitem [{\citenamefont {N\'eel}\ \emph {et~al.}(2010)\citenamefont {N\'eel},
  \citenamefont {Kr\"oger},\ and\ \citenamefont {Berndt}}]{prb_82_233401}%
  \BibitemOpen
  \bibfield  {author} {\bibinfo {author} {\bibfnamefont {N.}~\bibnamefont
  {N\'eel}}, \bibinfo {author} {\bibfnamefont {J.}~\bibnamefont {Kr\"oger}}, \
  and\ \bibinfo {author} {\bibfnamefont {R.}~\bibnamefont {Berndt}},\ }\href
  {\doibase 10.1103/PhysRevB.82.233401} {\bibfield  {journal} {\bibinfo
  {journal} {Phys. Rev. B}\ }\textbf {\bibinfo {volume} {82}},\ \bibinfo
  {pages} {233401} (\bibinfo {year} {2010})}\BibitemShut {NoStop}%
\bibitem [{\citenamefont {Kügel}\ \emph {et~al.}(2014)\citenamefont {Kügel},
  \citenamefont {Karolak}, \citenamefont {Senkpiel}, \citenamefont {Hsu},
  \citenamefont {Sangiovanni},\ and\ \citenamefont {Bode}}]{nl_14_3895}%
  \BibitemOpen
  \bibfield  {author} {\bibinfo {author} {\bibfnamefont {J.}~\bibnamefont
  {Kügel}}, \bibinfo {author} {\bibfnamefont {M.}~\bibnamefont {Karolak}},
  \bibinfo {author} {\bibfnamefont {J.}~\bibnamefont {Senkpiel}}, \bibinfo
  {author} {\bibfnamefont {P.-J.}\ \bibnamefont {Hsu}}, \bibinfo {author}
  {\bibfnamefont {G.}~\bibnamefont {Sangiovanni}}, \ and\ \bibinfo {author}
  {\bibfnamefont {M.}~\bibnamefont {Bode}},\ }\href {\doibase
  10.1021/nl501150k} {\bibfield  {journal} {\bibinfo  {journal} {Nano Letters}\
  }\textbf {\bibinfo {volume} {14}},\ \bibinfo {pages} {3895} (\bibinfo {year}
  {2014})}\BibitemShut {NoStop}%
\bibitem [{\citenamefont {Knaak}\ \emph {et~al.}(2017)\citenamefont {Knaak},
  \citenamefont {Gruber}, \citenamefont {Lindström}, \citenamefont {Bocquet},
  \citenamefont {Heck},\ and\ \citenamefont {Berndt}}]{nl_17_7146}%
  \BibitemOpen
  \bibfield  {author} {\bibinfo {author} {\bibfnamefont {T.}~\bibnamefont
  {Knaak}}, \bibinfo {author} {\bibfnamefont {M.}~\bibnamefont {Gruber}},
  \bibinfo {author} {\bibfnamefont {C.}~\bibnamefont {Lindström}}, \bibinfo
  {author} {\bibfnamefont {M.-L.}\ \bibnamefont {Bocquet}}, \bibinfo {author}
  {\bibfnamefont {J.}~\bibnamefont {Heck}}, \ and\ \bibinfo {author}
  {\bibfnamefont {R.}~\bibnamefont {Berndt}},\ }\href {\doibase
  10.1021/acs.nanolett.7b04181} {\bibfield  {journal} {\bibinfo  {journal}
  {Nano Letters}\ }\textbf {\bibinfo {volume} {17}},\ \bibinfo {pages} {7146}
  (\bibinfo {year} {2017})}\BibitemShut {NoStop}%
\bibitem [{\citenamefont {Limot}\ and\ \citenamefont
  {Berndt}(2004)}]{ass_237_576}%
  \BibitemOpen
  \bibfield  {author} {\bibinfo {author} {\bibfnamefont {L.}~\bibnamefont
  {Limot}}\ and\ \bibinfo {author} {\bibfnamefont {R.}~\bibnamefont {Berndt}},\
  }\href {\doibase https://doi.org/10.1016/j.apsusc.2004.07.023} {\bibfield
  {journal} {\bibinfo  {journal} {Applied Surface Science}\ }\textbf {\bibinfo
  {volume} {237}},\ \bibinfo {pages} {572 } (\bibinfo {year}
  {2004})}\BibitemShut {NoStop}%
\bibitem [{\citenamefont {Uchihashi}\ \emph {et~al.}(2008)\citenamefont
  {Uchihashi}, \citenamefont {Zhang}, \citenamefont {Kr\"oger},\ and\
  \citenamefont {Berndt}}]{prb_78_033402}%
  \BibitemOpen
  \bibfield  {author} {\bibinfo {author} {\bibfnamefont {T.}~\bibnamefont
  {Uchihashi}}, \bibinfo {author} {\bibfnamefont {J.}~\bibnamefont {Zhang}},
  \bibinfo {author} {\bibfnamefont {J.}~\bibnamefont {Kr\"oger}}, \ and\
  \bibinfo {author} {\bibfnamefont {R.}~\bibnamefont {Berndt}},\ }\href
  {\doibase 10.1103/PhysRevB.78.033402} {\bibfield  {journal} {\bibinfo
  {journal} {Phys. Rev. B}\ }\textbf {\bibinfo {volume} {78}},\ \bibinfo
  {pages} {033402} (\bibinfo {year} {2008})}\BibitemShut {NoStop}%
\bibitem [{\citenamefont {Li}\ \emph {et~al.}(2018)\citenamefont {Li},
  \citenamefont {Zheng}, \citenamefont {Wang}, \citenamefont {Miao},
  \citenamefont {Cao}, \citenamefont {Sun}, \citenamefont {Wu}, \citenamefont
  {Wu}, \citenamefont {Li}, \citenamefont {Wang},\ and\ \citenamefont
  {Ding}}]{prb_97_035417}%
  \BibitemOpen
  \bibfield  {author} {\bibinfo {author} {\bibfnamefont {Q.~L.}\ \bibnamefont
  {Li}}, \bibinfo {author} {\bibfnamefont {C.}~\bibnamefont {Zheng}}, \bibinfo
  {author} {\bibfnamefont {R.}~\bibnamefont {Wang}}, \bibinfo {author}
  {\bibfnamefont {B.~F.}\ \bibnamefont {Miao}}, \bibinfo {author}
  {\bibfnamefont {R.~X.}\ \bibnamefont {Cao}}, \bibinfo {author} {\bibfnamefont
  {L.}~\bibnamefont {Sun}}, \bibinfo {author} {\bibfnamefont {D.}~\bibnamefont
  {Wu}}, \bibinfo {author} {\bibfnamefont {Y.~Z.}\ \bibnamefont {Wu}}, \bibinfo
  {author} {\bibfnamefont {S.~C.}\ \bibnamefont {Li}}, \bibinfo {author}
  {\bibfnamefont {B.~G.}\ \bibnamefont {Wang}}, \ and\ \bibinfo {author}
  {\bibfnamefont {H.~F.}\ \bibnamefont {Ding}},\ }\href {\doibase
  10.1103/PhysRevB.97.035417} {\bibfield  {journal} {\bibinfo  {journal} {Phys.
  Rev. B}\ }\textbf {\bibinfo {volume} {97}},\ \bibinfo {pages} {035417}
  (\bibinfo {year} {2018})}\BibitemShut {NoStop}%
\bibitem [{\citenamefont {Vollhardt}\ and\ \citenamefont
  {Lichtenstein}(2017)}]{epjst_226_2439}%
  \BibitemOpen
  \bibfield  {author} {\bibinfo {author} {\bibfnamefont {D.}~\bibnamefont
  {Vollhardt}}\ and\ \bibinfo {author} {\bibfnamefont {A.~I.}\ \bibnamefont
  {Lichtenstein}},\ }\href {\doibase 10.1140/epjst/e2017-70078-x} {\bibfield
  {journal} {\bibinfo  {journal} {Eur. Phys. J. Spec. Top.}\ }\textbf {\bibinfo
  {volume} {226}},\ \bibinfo {pages} {2439} (\bibinfo {year}
  {2017})}\BibitemShut {NoStop}%
\bibitem [{\citenamefont {Limot}\ \emph {et~al.}(2005)\citenamefont {Limot},
  \citenamefont {Kr\"oger}, \citenamefont {Berndt}, \citenamefont
  {Garcia-Lekue},\ and\ \citenamefont {Hofer}}]{prl_94_126102}%
  \BibitemOpen
  \bibfield  {author} {\bibinfo {author} {\bibfnamefont {L.}~\bibnamefont
  {Limot}}, \bibinfo {author} {\bibfnamefont {J.}~\bibnamefont {Kr\"oger}},
  \bibinfo {author} {\bibfnamefont {R.}~\bibnamefont {Berndt}}, \bibinfo
  {author} {\bibfnamefont {A.}~\bibnamefont {Garcia-Lekue}}, \ and\ \bibinfo
  {author} {\bibfnamefont {W.~A.}\ \bibnamefont {Hofer}},\ }\href {\doibase
  10.1103/PhysRevLett.94.126102} {\bibfield  {journal} {\bibinfo  {journal}
  {Phys. Rev. Lett.}\ }\textbf {\bibinfo {volume} {94}},\ \bibinfo {pages}
  {126102} (\bibinfo {year} {2005})}\BibitemShut {NoStop}%
\bibitem [{\citenamefont {Kr\"oger}\ \emph {et~al.}(2007)\citenamefont
  {Kr\"oger}, \citenamefont {Jensen},\ and\ \citenamefont
  {Berndt}}]{njp_9_153}%
  \BibitemOpen
  \bibfield  {author} {\bibinfo {author} {\bibfnamefont {J.}~\bibnamefont
  {Kr\"oger}}, \bibinfo {author} {\bibfnamefont {H.}~\bibnamefont {Jensen}}, \
  and\ \bibinfo {author} {\bibfnamefont {R.}~\bibnamefont {Berndt}},\ }\href
  {http://stacks.iop.org/1367-2630/9/i=5/a=153} {\bibfield  {journal} {\bibinfo
   {journal} {New J. Phys.}\ }\textbf {\bibinfo {volume} {9}},\ \bibinfo
  {pages} {153} (\bibinfo {year} {2007})}\BibitemShut {NoStop}%
\bibitem [{\citenamefont {Kr\"oger}\ \emph {et~al.}(2008)\citenamefont
  {Kr\"oger}, \citenamefont {N\'eel},\ and\ \citenamefont
  {Limot}}]{jpcm_20_223001}%
  \BibitemOpen
  \bibfield  {author} {\bibinfo {author} {\bibfnamefont {J.}~\bibnamefont
  {Kr\"oger}}, \bibinfo {author} {\bibfnamefont {N.}~\bibnamefont {N\'eel}}, \
  and\ \bibinfo {author} {\bibfnamefont {L.}~\bibnamefont {Limot}},\ }\href
  {http://stacks.iop.org/0953-8984/20/i=22/a=223001} {\bibfield  {journal}
  {\bibinfo  {journal} {J. Phys.: Condens. Matter}\ }\textbf {\bibinfo {volume}
  {20}},\ \bibinfo {pages} {223001} (\bibinfo {year} {2008})}\BibitemShut
  {NoStop}%
\bibitem [{\citenamefont {Kr\"oger}\ \emph {et~al.}(2009)\citenamefont
  {Kr\"oger}, \citenamefont {N\'eel}, \citenamefont {Sperl}, \citenamefont
  {Wang},\ and\ \citenamefont {Berndt}}]{njp_11_125006}%
  \BibitemOpen
  \bibfield  {author} {\bibinfo {author} {\bibfnamefont {J.}~\bibnamefont
  {Kr\"oger}}, \bibinfo {author} {\bibfnamefont {N.}~\bibnamefont {N\'eel}},
  \bibinfo {author} {\bibfnamefont {A.}~\bibnamefont {Sperl}}, \bibinfo
  {author} {\bibfnamefont {Y.~F.}\ \bibnamefont {Wang}}, \ and\ \bibinfo
  {author} {\bibfnamefont {R.}~\bibnamefont {Berndt}},\ }\href
  {http://stacks.iop.org/1367-2630/11/i=12/a=125006} {\bibfield  {journal}
  {\bibinfo  {journal} {New J. Phys.}\ }\textbf {\bibinfo {volume} {11}},\
  \bibinfo {pages} {125006} (\bibinfo {year} {2009})}\BibitemShut {NoStop}%
\bibitem [{\citenamefont {Horcas}\ \emph {et~al.}(2007)\citenamefont {Horcas},
  \citenamefont {Fern{\'{a}}ndez}, \citenamefont {G{\'{o}}mez-Rodr{\'{i}}guez},
  \citenamefont {Colchero}, \citenamefont {G{\'{o}}mez-Herrero},\ and\
  \citenamefont {Baro}}]{rsi_78_013705}%
  \BibitemOpen
  \bibfield  {author} {\bibinfo {author} {\bibfnamefont {I.}~\bibnamefont
  {Horcas}}, \bibinfo {author} {\bibfnamefont {R.}~\bibnamefont
  {Fern{\'{a}}ndez}}, \bibinfo {author} {\bibfnamefont {J.~M.}\ \bibnamefont
  {G{\'{o}}mez-Rodr{\'{i}}guez}}, \bibinfo {author} {\bibfnamefont
  {J.}~\bibnamefont {Colchero}}, \bibinfo {author} {\bibfnamefont
  {J.}~\bibnamefont {G{\'{o}}mez-Herrero}}, \ and\ \bibinfo {author}
  {\bibfnamefont {A.~M.}\ \bibnamefont {Baro}},\ }\href {\doibase
  10.1063/1.2432410} {\bibfield  {journal} {\bibinfo  {journal} {Rev. Sci.
  Instrum.}\ }\textbf {\bibinfo {volume} {78}},\ \bibinfo {pages} {013705}
  (\bibinfo {year} {2007})}\BibitemShut {NoStop}%
\bibitem [{\citenamefont {Stroscio}\ \emph {et~al.}(2006)\citenamefont
  {Stroscio}, \citenamefont {Tavazza}, \citenamefont {Crain}, \citenamefont
  {Celotta},\ and\ \citenamefont {Chaka}}]{science_313_948}%
  \BibitemOpen
  \bibfield  {author} {\bibinfo {author} {\bibfnamefont {J.~A.}\ \bibnamefont
  {Stroscio}}, \bibinfo {author} {\bibfnamefont {F.}~\bibnamefont {Tavazza}},
  \bibinfo {author} {\bibfnamefont {J.~N.}\ \bibnamefont {Crain}}, \bibinfo
  {author} {\bibfnamefont {R.~J.}\ \bibnamefont {Celotta}}, \ and\ \bibinfo
  {author} {\bibfnamefont {A.~M.}\ \bibnamefont {Chaka}},\ }\href {\doibase
  10.1126/science.1129788} {\bibfield  {journal} {\bibinfo  {journal}
  {Science}\ }\textbf {\bibinfo {volume} {313}},\ \bibinfo {pages} {948}
  (\bibinfo {year} {2006})}\BibitemShut {NoStop}%
\bibitem [{\citenamefont {Lagoute}\ \emph {et~al.}(2007)\citenamefont
  {Lagoute}, \citenamefont {Nacci},\ and\ \citenamefont
  {F\"olsch}}]{prl_98_146804}%
  \BibitemOpen
  \bibfield  {author} {\bibinfo {author} {\bibfnamefont {J.}~\bibnamefont
  {Lagoute}}, \bibinfo {author} {\bibfnamefont {C.}~\bibnamefont {Nacci}}, \
  and\ \bibinfo {author} {\bibfnamefont {S.}~\bibnamefont {F\"olsch}},\ }\href
  {\doibase 10.1103/PhysRevLett.98.146804} {\bibfield  {journal} {\bibinfo
  {journal} {Phys. Rev. Lett.}\ }\textbf {\bibinfo {volume} {98}},\ \bibinfo
  {pages} {146804} (\bibinfo {year} {2007})}\BibitemShut {NoStop}%
\bibitem [{\citenamefont {Kresse}\ and\ \citenamefont
  {Hafner}(1994)}]{jpcm_6_8245}%
  \BibitemOpen
  \bibfield  {author} {\bibinfo {author} {\bibfnamefont {G.}~\bibnamefont
  {Kresse}}\ and\ \bibinfo {author} {\bibfnamefont {J.}~\bibnamefont
  {Hafner}},\ }\href {\doibase 10.1088/0953-8984/6/40/015} {\bibfield
  {journal} {\bibinfo  {journal} {J. Phys.: Condens. Matter}\ }\textbf
  {\bibinfo {volume} {6}},\ \bibinfo {pages} {8245} (\bibinfo {year}
  {1994})}\BibitemShut {NoStop}%
\bibitem [{\citenamefont {Bl\"ochl}(1994)}]{prb_50_17953}%
  \BibitemOpen
  \bibfield  {author} {\bibinfo {author} {\bibfnamefont {P.~E.}\ \bibnamefont
  {Bl\"ochl}},\ }\href {\doibase 10.1103/PhysRevB.50.17953} {\bibfield
  {journal} {\bibinfo  {journal} {Phys. Rev. B}\ }\textbf {\bibinfo {volume}
  {50}},\ \bibinfo {pages} {17953} (\bibinfo {year} {1994})}\BibitemShut
  {NoStop}%
\bibitem [{\citenamefont {Kresse}\ and\ \citenamefont
  {Joubert}(1999)}]{prb_59_1758}%
  \BibitemOpen
  \bibfield  {author} {\bibinfo {author} {\bibfnamefont {G.}~\bibnamefont
  {Kresse}}\ and\ \bibinfo {author} {\bibfnamefont {D.}~\bibnamefont
  {Joubert}},\ }\href {\doibase 10.1103/PhysRevB.59.1758} {\bibfield  {journal}
  {\bibinfo  {journal} {Phys. Rev. B}\ }\textbf {\bibinfo {volume} {59}},\
  \bibinfo {pages} {1758} (\bibinfo {year} {1999})}\BibitemShut {NoStop}%
\bibitem [{\citenamefont {Perdew}\ \emph {et~al.}(1996)\citenamefont {Perdew},
  \citenamefont {Burke},\ and\ \citenamefont {Ernzerhof}}]{prl_77_3865}%
  \BibitemOpen
  \bibfield  {author} {\bibinfo {author} {\bibfnamefont {J.~P.}\ \bibnamefont
  {Perdew}}, \bibinfo {author} {\bibfnamefont {K.}~\bibnamefont {Burke}}, \
  and\ \bibinfo {author} {\bibfnamefont {M.}~\bibnamefont {Ernzerhof}},\ }\href
  {\doibase 10.1103/PhysRevLett.77.3865} {\bibfield  {journal} {\bibinfo
  {journal} {Phys. Rev. Lett.}\ }\textbf {\bibinfo {volume} {77}},\ \bibinfo
  {pages} {3865} (\bibinfo {year} {1996})}\BibitemShut {NoStop}%
\bibitem [{\citenamefont {Amadon}\ \emph {et~al.}(2008)\citenamefont {Amadon},
  \citenamefont {Lechermann}, \citenamefont {Georges}, \citenamefont {Jollet},
  \citenamefont {Wehling},\ and\ \citenamefont {Lichtenstein}}]{prb_77_205112}%
  \BibitemOpen
  \bibfield  {author} {\bibinfo {author} {\bibfnamefont {B.}~\bibnamefont
  {Amadon}}, \bibinfo {author} {\bibfnamefont {F.}~\bibnamefont {Lechermann}},
  \bibinfo {author} {\bibfnamefont {A.}~\bibnamefont {Georges}}, \bibinfo
  {author} {\bibfnamefont {F.}~\bibnamefont {Jollet}}, \bibinfo {author}
  {\bibfnamefont {T.~O.}\ \bibnamefont {Wehling}}, \ and\ \bibinfo {author}
  {\bibfnamefont {A.~I.}\ \bibnamefont {Lichtenstein}},\ }\href {\doibase
  10.1103/PhysRevB.77.205112} {\bibfield  {journal} {\bibinfo  {journal} {Phys.
  Rev. B}\ }\textbf {\bibinfo {volume} {77}},\ \bibinfo {pages} {205112}
  (\bibinfo {year} {2008})}\BibitemShut {NoStop}%
\bibitem [{\citenamefont {Karolak}\ \emph {et~al.}(2011)\citenamefont
  {Karolak}, \citenamefont {Wehling}, \citenamefont {Lechermann},\ and\
  \citenamefont {Lichtenstein}}]{jpcm_23_085601}%
  \BibitemOpen
  \bibfield  {author} {\bibinfo {author} {\bibfnamefont {M.}~\bibnamefont
  {Karolak}}, \bibinfo {author} {\bibfnamefont {T.~O.}\ \bibnamefont
  {Wehling}}, \bibinfo {author} {\bibfnamefont {F.}~\bibnamefont {Lechermann}},
  \ and\ \bibinfo {author} {\bibfnamefont {A.~I.}\ \bibnamefont
  {Lichtenstein}},\ }\href {\doibase 10.1088/0953-8984/23/8/085601} {\bibfield
  {journal} {\bibinfo  {journal} {Journal of Physics: Condensed Matter}\
  }\textbf {\bibinfo {volume} {23}},\ \bibinfo {pages} {085601} (\bibinfo
  {year} {2011})}\BibitemShut {NoStop}%
\bibitem [{\citenamefont {Surer}\ \emph {et~al.}(2012)\citenamefont {Surer},
  \citenamefont {Troyer}, \citenamefont {Werner}, \citenamefont {Wehling},
  \citenamefont {L\"auchli}, \citenamefont {Wilhelm},\ and\ \citenamefont
  {Lichtenstein}}]{prb_85_085114}%
  \BibitemOpen
  \bibfield  {author} {\bibinfo {author} {\bibfnamefont {B.}~\bibnamefont
  {Surer}}, \bibinfo {author} {\bibfnamefont {M.}~\bibnamefont {Troyer}},
  \bibinfo {author} {\bibfnamefont {P.}~\bibnamefont {Werner}}, \bibinfo
  {author} {\bibfnamefont {T.~O.}\ \bibnamefont {Wehling}}, \bibinfo {author}
  {\bibfnamefont {A.~M.}\ \bibnamefont {L\"auchli}}, \bibinfo {author}
  {\bibfnamefont {A.}~\bibnamefont {Wilhelm}}, \ and\ \bibinfo {author}
  {\bibfnamefont {A.~I.}\ \bibnamefont {Lichtenstein}},\ }\href {\doibase
  10.1103/PhysRevB.85.085114} {\bibfield  {journal} {\bibinfo  {journal} {Phys.
  Rev. B}\ }\textbf {\bibinfo {volume} {85}},\ \bibinfo {pages} {085114}
  (\bibinfo {year} {2012})}\BibitemShut {NoStop}%
\bibitem [{\citenamefont {{Nozi\`eres, Ph.}}\ and\ \citenamefont {{Blandin,
  A.}}(1980)}]{jp_41_193}%
  \BibitemOpen
  \bibfield  {author} {\bibinfo {author} {\bibnamefont {{Nozi\`eres, Ph.}}}\
  and\ \bibinfo {author} {\bibnamefont {{Blandin, A.}}},\ }\href {\doibase
  10.1051/jphys:01980004103019300} {\bibfield  {journal} {\bibinfo  {journal}
  {J. Phys. France}\ }\textbf {\bibinfo {volume} {41}},\ \bibinfo {pages} {193}
  (\bibinfo {year} {1980})}\BibitemShut {NoStop}%
\bibitem [{\citenamefont {Hewson}(1993)}]{cup_1993}%
  \BibitemOpen
  \bibfield  {author} {\bibinfo {author} {\bibfnamefont {A.~C.}\ \bibnamefont
  {Hewson}},\ }\href@noop {} {\emph {\bibinfo {title} {The Kondo Problem to
  Heavy Fermions}}}\ (\bibinfo  {publisher} {Cambridge University Press},\
  \bibinfo {year} {1993})\BibitemShut {NoStop}%
\bibitem [{\citenamefont {Werner}\ \emph {et~al.}(2006)\citenamefont {Werner},
  \citenamefont {Comanac}, \citenamefont {de' Medici}, \citenamefont {Troyer},\
  and\ \citenamefont {Millis}}]{prl_97_076405}%
  \BibitemOpen
  \bibfield  {author} {\bibinfo {author} {\bibfnamefont {P.}~\bibnamefont
  {Werner}}, \bibinfo {author} {\bibfnamefont {A.}~\bibnamefont {Comanac}},
  \bibinfo {author} {\bibfnamefont {L.}~\bibnamefont {de' Medici}}, \bibinfo
  {author} {\bibfnamefont {M.}~\bibnamefont {Troyer}}, \ and\ \bibinfo {author}
  {\bibfnamefont {A.~J.}\ \bibnamefont {Millis}},\ }\href {\doibase
  10.1103/PhysRevLett.97.076405} {\bibfield  {journal} {\bibinfo  {journal}
  {Phys. Rev. Lett.}\ }\textbf {\bibinfo {volume} {97}},\ \bibinfo {pages}
  {076405} (\bibinfo {year} {2006})}\BibitemShut {NoStop}%
\bibitem [{\citenamefont {Werner}\ and\ \citenamefont
  {Millis}(2006)}]{prb_74_155107}%
  \BibitemOpen
  \bibfield  {author} {\bibinfo {author} {\bibfnamefont {P.}~\bibnamefont
  {Werner}}\ and\ \bibinfo {author} {\bibfnamefont {A.~J.}\ \bibnamefont
  {Millis}},\ }\href {\doibase 10.1103/PhysRevB.74.155107} {\bibfield
  {journal} {\bibinfo  {journal} {Phys. Rev. B}\ }\textbf {\bibinfo {volume}
  {74}},\ \bibinfo {pages} {155107} (\bibinfo {year} {2006})}\BibitemShut
  {NoStop}%
\bibitem [{\citenamefont {Wallerberger}\ \emph {et~al.}(2019)\citenamefont
  {Wallerberger}, \citenamefont {Hausoel}, \citenamefont {Gunacker},
  \citenamefont {Kowalski}, \citenamefont {Parragh}, \citenamefont {Goth},
  \citenamefont {Held},\ and\ \citenamefont {Sangiovanni}}]{cpc_235_388}%
  \BibitemOpen
  \bibfield  {author} {\bibinfo {author} {\bibfnamefont {M.}~\bibnamefont
  {Wallerberger}}, \bibinfo {author} {\bibfnamefont {A.}~\bibnamefont
  {Hausoel}}, \bibinfo {author} {\bibfnamefont {P.}~\bibnamefont {Gunacker}},
  \bibinfo {author} {\bibfnamefont {A.}~\bibnamefont {Kowalski}}, \bibinfo
  {author} {\bibfnamefont {N.}~\bibnamefont {Parragh}}, \bibinfo {author}
  {\bibfnamefont {F.}~\bibnamefont {Goth}}, \bibinfo {author} {\bibfnamefont
  {K.}~\bibnamefont {Held}}, \ and\ \bibinfo {author} {\bibfnamefont
  {G.}~\bibnamefont {Sangiovanni}},\ }\href {\doibase
  https://doi.org/10.1016/j.cpc.2018.09.007} {\bibfield  {journal} {\bibinfo
  {journal} {Computer Physics Communications}\ }\textbf {\bibinfo {volume}
  {235}},\ \bibinfo {pages} {388 } (\bibinfo {year} {2019})}\BibitemShut
  {NoStop}%
\bibitem [{\citenamefont {\ifmmode \mbox{\c{S}}\else \c{S}\fi{}a\ifmmode
  \mbox{\c{s}}\else \c{s}\fi{}\ifmmode \imath \else \i
  \fi{}o\ifmmode~\breve{g}\else \u{g}\fi{}lu}\ \emph
  {et~al.}(2011)\citenamefont {\ifmmode \mbox{\c{S}}\else \c{S}\fi{}a\ifmmode
  \mbox{\c{s}}\else \c{s}\fi{}\ifmmode \imath \else \i
  \fi{}o\ifmmode~\breve{g}\else \u{g}\fi{}lu}, \citenamefont {Friedrich},\ and\
  \citenamefont {Bl\"ugel}}]{prb_83_121101}%
  \BibitemOpen
  \bibfield  {author} {\bibinfo {author} {\bibfnamefont {E.}~\bibnamefont
  {\ifmmode \mbox{\c{S}}\else \c{S}\fi{}a\ifmmode \mbox{\c{s}}\else
  \c{s}\fi{}\ifmmode \imath \else \i \fi{}o\ifmmode~\breve{g}\else
  \u{g}\fi{}lu}}, \bibinfo {author} {\bibfnamefont {C.}~\bibnamefont
  {Friedrich}}, \ and\ \bibinfo {author} {\bibfnamefont {S.}~\bibnamefont
  {Bl\"ugel}},\ }\href {\doibase 10.1103/PhysRevB.83.121101} {\bibfield
  {journal} {\bibinfo  {journal} {Phys. Rev. B}\ }\textbf {\bibinfo {volume}
  {83}},\ \bibinfo {pages} {121101} (\bibinfo {year} {2011})}\BibitemShut
  {NoStop}%
\bibitem [{\citenamefont {\ifmmode \mbox{\c{S}}\else \c{S}\fi{}a\ifmmode
  \mbox{\c{s}}\else \c{s}\fi{}\ifmmode \imath \else \i
  \fi{}o\ifmmode~\breve{g}\else \u{g}\fi{}lu}\ \emph
  {et~al.}(2012)\citenamefont {\ifmmode \mbox{\c{S}}\else \c{S}\fi{}a\ifmmode
  \mbox{\c{s}}\else \c{s}\fi{}\ifmmode \imath \else \i
  \fi{}o\ifmmode~\breve{g}\else \u{g}\fi{}lu}, \citenamefont {Friedrich},\ and\
  \citenamefont {Bl\"ugel}}]{prl_109_146401}%
  \BibitemOpen
  \bibfield  {author} {\bibinfo {author} {\bibfnamefont {E.}~\bibnamefont
  {\ifmmode \mbox{\c{S}}\else \c{S}\fi{}a\ifmmode \mbox{\c{s}}\else
  \c{s}\fi{}\ifmmode \imath \else \i \fi{}o\ifmmode~\breve{g}\else
  \u{g}\fi{}lu}}, \bibinfo {author} {\bibfnamefont {C.}~\bibnamefont
  {Friedrich}}, \ and\ \bibinfo {author} {\bibfnamefont {S.}~\bibnamefont
  {Bl\"ugel}},\ }\href {\doibase 10.1103/PhysRevLett.109.146401} {\bibfield
  {journal} {\bibinfo  {journal} {Phys. Rev. Lett.}\ }\textbf {\bibinfo
  {volume} {109}},\ \bibinfo {pages} {146401} (\bibinfo {year}
  {2012})}\BibitemShut {NoStop}%
\bibitem [{\citenamefont {Chubukov}\ and\ \citenamefont
  {Maslov}(2012)}]{prb_86_155136}%
  \BibitemOpen
  \bibfield  {author} {\bibinfo {author} {\bibfnamefont {A.~V.}\ \bibnamefont
  {Chubukov}}\ and\ \bibinfo {author} {\bibfnamefont {D.~L.}\ \bibnamefont
  {Maslov}},\ }\href {\doibase 10.1103/PhysRevB.86.155136} {\bibfield
  {journal} {\bibinfo  {journal} {Phys. Rev. B}\ }\textbf {\bibinfo {volume}
  {86}},\ \bibinfo {pages} {155136} (\bibinfo {year} {2012})}\BibitemShut
  {NoStop}%
\bibitem [{\citenamefont {Jarrell}\ and\ \citenamefont
  {Gubernatis}(1996)}]{pr_269_133}%
  \BibitemOpen
  \bibfield  {author} {\bibinfo {author} {\bibfnamefont {M.}~\bibnamefont
  {Jarrell}}\ and\ \bibinfo {author} {\bibfnamefont {J.}~\bibnamefont
  {Gubernatis}},\ }\href {\doibase
  https://doi.org/10.1016/0370-1573(95)00074-7} {\bibfield  {journal} {\bibinfo
   {journal} {Physics Reports}\ }\textbf {\bibinfo {volume} {269}},\ \bibinfo
  {pages} {133 } (\bibinfo {year} {1996})}\BibitemShut {NoStop}%
\bibitem [{\citenamefont {Bryan}(1990)}]{ebj_18_165}%
  \BibitemOpen
  \bibfield  {author} {\bibinfo {author} {\bibfnamefont {R.~K.}\ \bibnamefont
  {Bryan}},\ }\href {\doibase 10.1007/BF02427376} {\bibfield  {journal}
  {\bibinfo  {journal} {European Biophysics Journal}\ }\textbf {\bibinfo
  {volume} {18}},\ \bibinfo {pages} {165} (\bibinfo {year} {1990})}\BibitemShut
  {NoStop}%
\bibitem [{\citenamefont {Levy}\ \emph {et~al.}(2017)\citenamefont {Levy},
  \citenamefont {LeBlanc},\ and\ \citenamefont {Gull}}]{cpc_215_149}%
  \BibitemOpen
  \bibfield  {author} {\bibinfo {author} {\bibfnamefont {R.}~\bibnamefont
  {Levy}}, \bibinfo {author} {\bibfnamefont {J.}~\bibnamefont {LeBlanc}}, \
  and\ \bibinfo {author} {\bibfnamefont {E.}~\bibnamefont {Gull}},\ }\href
  {\doibase https://doi.org/10.1016/j.cpc.2017.01.018} {\bibfield  {journal}
  {\bibinfo  {journal} {Computer Physics Communications}\ }\textbf {\bibinfo
  {volume} {215}},\ \bibinfo {pages} {149 } (\bibinfo {year}
  {2017})}\BibitemShut {NoStop}%
\bibitem [{\citenamefont {C.~Hewson}(2005)}]{jpsj_74_8}%
  \BibitemOpen
  \bibfield  {author} {\bibinfo {author} {\bibfnamefont {A.}~\bibnamefont
  {C.~Hewson}},\ }\href {\doibase 10.1143/JPSJ.74.8} {\bibfield  {journal}
  {\bibinfo  {journal} {Journal of the Physical Society of Japan}\ }\textbf
  {\bibinfo {volume} {74}},\ \bibinfo {pages} {8} (\bibinfo {year}
  {2005})}\BibitemShut {NoStop}%
\bibitem [{\citenamefont {Edwards}\ and\ \citenamefont
  {Hewson}(2011)}]{jpcm_23_045601}%
  \BibitemOpen
  \bibfield  {author} {\bibinfo {author} {\bibfnamefont {K.}~\bibnamefont
  {Edwards}}\ and\ \bibinfo {author} {\bibfnamefont {A.~C.}\ \bibnamefont
  {Hewson}},\ }\href {\doibase 10.1088/0953-8984/23/4/045601} {\bibfield
  {journal} {\bibinfo  {journal} {Journal of Physics: Condensed Matter}\
  }\textbf {\bibinfo {volume} {23}},\ \bibinfo {pages} {045601} (\bibinfo
  {year} {2011})}\BibitemShut {NoStop}%
\end{thebibliography}
%

\end{document}